\documentclass{article}

\RequirePackage{amsthm,amsmath,amsfonts,amssymb}
\RequirePackage[authoryear]{natbib}

\usepackage[a4paper]{geometry}

\usepackage{graphicx} 
\usepackage{amsmath}
\usepackage{tikz}
\usetikzlibrary{arrows.meta,
                decorations.markings}
\usepackage{mathrsfs}
\usepackage{amsfonts}
\usepackage{natbib}
\usepackage{url}
\usetikzlibrary{arrows}
\usepackage{tipa}

\newcommand{\e}{\mathrm{e}}

\usepackage{xspace}

\newcommand{\ourmethod}{WaST\xspace}

\newcommand{\locus}{\emph{node}\xspace}
\newcommand{\loci}{\emph{nodes}\xspace}

\title{\ourmethod: a formalisation of the Wave model with associated statistical inference and applications}

\author{Grégoire Clarté, University of Edinburgh, School of Maths}

\begin{document}

\maketitle

\begin{abstract}
    We propose a mathematical formalisation of the ``wave model'' originally developed in historical linguistics but with further applications in human sciences. This model assumes new traits appear in a population and spread to nearby populations depending on their closeness. It is mostly used to describe joint evolution of closely related populations, for example of several dialects. These situations of permanent contact are not accurately represented by its competitors based on tree structures. We built a fully Bayesian generative model where innovation spread along a fixed graph and disappear according to a death process. We then  develop a Metropolis-Hastings within Gibbs sampler to sample from the posterior distribution on the graph. We test our method on simulated datasets as well as on several real dataset.
\end{abstract}

\section{Introduction}

\citet{schmidt1872verwantschaftsverhaltnisse} proposed to represent the evolution of languages as innovations that appear in a population and spread to the nearby ones, like waves originating from different points in space. This concept, called \emph{Wellentheorie} and that we will call Wave Model, was one of the first description of language evolution, later applied to other contexts. In this description, it is  the superpositions of those waves create the observed diversity of the present languages \citep{trubetzkoy1939gedanken,aikhenvald2006areal}. This model strongly contrasts with the more studied Tree model, which assume no exchange between the populations.
If the model is sometimes considered as a ``mere conceptual metaphor'' \citep{pellard2024family}, it is mostly because of the lack of a proper formalisation that allows for discussion of the assumptions and implications of such a model. The Wave model assumes that there are different ``populations'' (we will discuss this notion in a later part) among which innovations spread. The spread occurs following geographical and social lines, but at any moment the spread can stop leaving some populations without the innovation. This model has the particularity of not assuming any dynamic on the populations themselves --- they don't merge or split ---, and thus can explain small changes and variations in groups of interacting groups of people.

Previous attempts at representing Waves of innovations often focused on observing the maximal state of expansion of innovations or to measure the distances between the languages by considering presence or absence of traits \citep{anttila1989historical,kalyan2018freeing,franccois2015trees,Pelkey2024Wave}. These works have allowed a more quantitative approach to the Wave Model, but did not propose a generative, and chronological, interpretation of the mode: there is no time factor in these descriptive works. Some other works \citep{Hartmann2023Wave} did not lead to statistical inference due to their complexity.

Here, we propose a generative description of the Wave model.  The general idea of our model is to represent the different populations and their degree of connection as a metric graph (i.e. a network). Innovations start at any vertex of the graph and spreads along the edges of the graph. An innovation can then disappear following a death process  either during the spread on the edges or at a node. 

Expressed as above, this model deals with binary observation (an innovation is either absent or present at a vertex). This model is particularly adapted to the study of lexical innovations in linguistics, but less to the study of grammatical innovations or cultural aspects. Further improvements would be needed for proper applications to those fields. This article will focus on applications to linguistics, from which the Wave model originates, and cultural studies. Among possible applications outside of linguistics, we can mention \citet{dolbunova2023transmission} in which the authors are interested in how ceramics techniques and decors have spread between the populations. The network on which the spread occurs is reconstituted with geographical information.

The problem of learning an unknown metric graph shares similarities with Gaussian Graphical Models  \citep{margolin2006aracne,hawe2019inferring}, where nodes represent variables and the length of the edges the correlations between them. Other similarities can be found in  the reconstitution of interaction network in epidemiology or in the study of social structures \citep{lloyd2007network}. The model we present differs in it being a diachronic generative model, that does not enjoy the simple Gaussian distribution properties.

In the future, this work will allow comparison with its main competitor in Linguistics, the Tree models, which assumes heredity is the main drive of evolution of languages. These models have numerous variations and applications \citep{lewis2001likelihood,tuffley1998modeling,Dollo1887,Nicholls2007dated,GrayRussellAtkinsonQuentin}. Most often, a Tree model is used because there is no quantitative alternative. Proposing a first implementation of the Wave model, even with its limited scope has been expected by some linguists \citep{Geisler2022}. Furthermore,  the type of data used for most Tree models, especially Dollo and Covarion, is the same as the model we present here. Making particularly simple the comparison of the two outputs. More details on the ongoing debates around Tree models and Wave models can be found in \citet{pellard2024family}.

After describing the datasets that can be used with the Wave model, we proceed by defining the model. We then develop a numerical method that allow to run inference on this model, and give several simulated and real data examples.

In this paper we will call indifferently the model and the numerical method of inference based on it \ourmethod, for Wave Spreading of Traits.

\section{Dataset}

The dataset is represented as a binary matrix, with each row corresponding to a trait (that is a trait), each column correspond to what we will call \locus, wich can be interpreted as a physical place or more broadly depending on the application. We discuss briefly this denomination in a later section. The value at the intersection represents whether or not the trait is present at the \locus. We assume all the columns are collected synchronously (that is there is no fossil). Note that this dataset has exactly the same specifications as the  dataset used with the Dollo and covarion models. In other words, any study run with Dollo model can be attempted with \ourmethod, with the limitations of  computational cost.

We represent in table \ref{tab:exampledataset} a suitable dataset from \citet{kalyan2019understanding} [personal communication]. Note that in this dataset several types of data are represented (Mrp: morphologic and Lex: lexical). The question on whether we can mix or not these traits is left for future works.

\begin{table}[]
    \centering
    \begin{tabular}{l|l|llllllll}
Type&Description&Hiw&Lo-Toga&Lehali&Löyöp&Volow&Mwotlap\\
\hline
Mrp& 1ex/2pl: *kam $\to$ $^{\text{\textipa{N}}}$gam- &1&1&1&1&1&1\\
Mrp& 3sg: *nia $>$ *gia&0&0&1&1&1&1\\
Lex& 
again: *\textipa{BaraG(a)i}??&1&1&1&0&0&0\\
Lex& 
again: /\textipa{sE}/&0&0&0&0&1&1\\
Lex& 
child: */\textipa{maGola}/&1&1&0&0&0&0\\
    \end{tabular}
    \caption{Example of binary dataset from the North Vanuatu dataset.}
    \label{tab:exampledataset}
\end{table}

We make the assumption that all the combinations of traits are possible, that is the presence of a trait cannot be constrained by the presence of another. In linguistics, this assumption is mostly true for lexical traits, way less for other types of traits.

The dataset is then $(D(k,\ell))_{k \in \{1,...,m\},\ell \in \{1,...,n\}}$, a matrix with values in $\{0,1\}$. With this notation there are $n$ \loci and $m$ traits to study.

\section{Generative model}

Given $V$ a set of $n$ \loci, we assume there exist a fixed (unknown) symmetric matrix $d = (d(\ell_1,\ell_2), \ell_1,\ell_2 \in V) \in \mathcal{S}_n(\mathbb{R^{+*}\cup \infty} )$ of (semi-)distances between the \loci that represents the connexion between the \loci. We expect this matrix to be ``sparse'' in that most of the elements should be $\infty$. By convention, we set $d(\ell,\ell)=\infty, \forall \ell \in V$. This is equivalent to defining a (undirected) metric  graph $\mathfrak{G}=(V,E,d)$, as with our definition of $d$, $E = \{ (\ell_1,\ell_2) \in V^2; d((\ell_1,\ell_2)) < \infty \}$.


Mathematically, we can write the process as, with unknown parameters $\mu,\nu,d,\mathfrak{G}$,

\begin{align}
    (t_i)_i &\sim \mathcal{P}oisson \mathcal{P}rocess(1); \label{eq:def1}\\
    \forall i, S_i &\sim^{iid} \mathcal{U}(V);  \label{eq:def2}\\
    \forall i, \mathcal{P}_i&\sim^{iid}  \bigotimes_{e \in E}\mathcal{B}(\exp(-\mu d_e));  \label{eq:def3}\\
    i.e. \ \forall i, \forall e \in E, P(e \in \mathcal{P}_i) &= \exp(-\mu d_e), \text{ independently} \nonumber\\
    \forall i, \forall j \in V, Z_j^i &\sim \mathcal{B}(\max(1,\exp(-\nu(t_i - d_{\mathcal{P}_i}(S_i,j) )); \label{eq:def4}\\
    \forall i, \forall j, D(i,j) &= Z^i_j \mathbf{1}_{d_{\mathcal{P}_i}(j,S_i) < t_i}. \label{eq:def5}
\end{align}

Where $d_{\mathcal{P}_i}(S_i,j)$ is the distance between $S_i$ and $j$ computed only through edges in $\mathcal{P}_i$: \[d_{\mathcal{P}_i}(S_i,j)= \underset{E \supset \mathcal{P}_i \supset c \text{ path from } S_i \text{ to } j}{\min} \vert c \vert_d,\]

with $c$ denoting a sequence of edges, and $\vert c \vert_d$ the sum of their length.

That is, with words: According to a Poisson process with rate $1$ (eq. \ref{eq:def1}), new traits appear at a random \locus, $S_i$, that is on a vertex of the graph (eq. \ref{eq:def2}). Then, the trait propagates at constant speed on all edges, but can die following a point process with rate $\mu$ independently on each edge. Equivalently, we can draw a Bernoulli distribution independently for each edge with a parameter that depends on its length to decide if they \emph{can} propagate the trait. This drawing occurs only once per trait: an edge that did not propagate a trait will never propagate it. In other words, we draw select a subset of edges by selecting independently each edge accoding to a probability that depends on its length (req. \ref{eq:def3}). We call \emph{pattern}, the selected set of edges, noted $\mathcal{P}_i$. Because of the independence of the other parts of the process, once this subset of edges is selected, this latent variable is used in the numerical part.

Once reaching the new \locus, it keeps propagating on the graph to all the connected \loci  \emph{that have not yet received it}. Each \locus that has the trait can lose it following a standard death process with rate $\nu$ (eq. \ref{eq:def5}). Thus, a trait at any \locus has a random survival time following $\mathcal{E}(\nu)$ starting from the time it acquired it (eq. \ref{eq:def4}). 

Note that we exclude the possibility that a disappeared trait spreads again to the same \loci. In other words, if a trait $i$ has several ways of spreading to a \locus (always given the same pattern $\mathcal{P}_i$), it will take the shortest path to the \locus and start disappearing after this point. 

We assume that the spread starts immediately once a trait has reached a \locus, and continues even if the trait later disappears. For example, we could imagine a trait spreading through a node that loses it immediately after gaining it, and still propagates it. 

\begin{figure}
    \centering
    \begin{tikzpicture}[
       decoration = {markings,
                     mark=at position .5 with {\arrow{Stealth[length=2mm]}}},
       dot/.style = {circle, fill, inner sep=2.4pt, node contents={},
                     label=#1}
                        ]
\node (a) at (0,5) [dot=$a$];
\node (b) at (1,4) [dot=below:$b$,red];
\node (e) at (4,2) [dot=right:$e$];
\node (g) at (6,4) [dot=right:$g$];
\node (f) at (3,0) [dot=right:$f$];
\node (c) at (3,3) [dot=right:$c$,blue];
\node (d) at (4,3.5) [dot=above:$d$,red];
\draw (a) edge (b) ;
\draw (c) edge (b) [red];
\draw (c) edge (d) [red];
\draw (e) edge (d) ;
\draw (d) edge (g) [red];
\draw (c) edge (e) ;
\draw (f) edge (e) [red];
    \draw[fill=none,red](c) circle (2.8);
    \end{tikzpicture}
    \caption{Example of spread of a trait. The circle represents the reachable nodes at the time of the observation, the red edges represent the edges on which it can spread, the red nodes where it is present, and the blue ones where it has disappeared. The associated row $D_{i,\cdot}$ in the dataset will then be $(0,1,0,1,0,0,0)$.}
    \label{fig:example_evolprocess}
\end{figure}
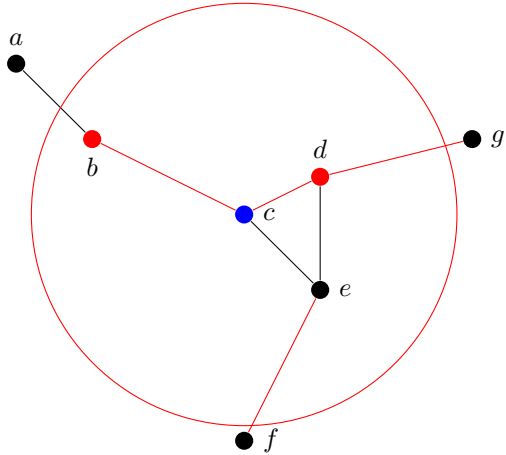

Figure \ref{fig:example_evolprocess} represents an example of evolution of a trait on the graph represented in Figure \ref{fig:example_underlying_graph}: the trait has source $c$, the circle represents the nodes reachable before observation, and the red edges represent  $\mathcal{P}_i$ --- reminder, the edges that \emph{could} spread the innovation---. Here, note that $g$, although connected by red edges to $c$, has not developed it before observation time, and that $f$ cannot develop it as there is no path of red (i.e. in $\mathcal{P}_i$) edges that lead to it. Furthermore, each node that develop it can lose it later, which is the case in $c$.

Note that the choice of a Poisson process with parameter $1$  for the apparition of the new traits allows the model to be identifiable. Unless time data is available (for example, date of apparition of a trait), this is the simplest way to make the model identifiable. This question is discussed in a later part.

\begin{figure}
    \centering
    \begin{tikzpicture}[
       decoration = {markings,
                     mark=at position .5 with {\arrow{Stealth[length=2mm]}}},
       dot/.style = {circle, fill, inner sep=2.4pt, node contents={},
                     label=#1}
                        ]
\node (a) at (0,5) [dot=$a$];
\node (b) at (1,4) [dot=below:$b$];
\node (e) at (4,2) [dot=right:$e$];
\node (g) at (6,4) [dot=right:$g$];
\node (f) at (3,0) [dot=right:$f$];
\node (c) at (3,3) [dot=right:$c$];
\node (d) at (4,3.5) [dot=above:$d$];
\draw (a) edge (b) ;
\draw (c) edge (b) ;
\draw (c) edge (d) ;
\draw (e) edge (d) ;
\draw (d) edge (g) ;
\draw (c) edge (e) ;
\draw (f) edge (e) ;
    \end{tikzpicture}
    \caption{Example of graph}
    \label{fig:example_underlying_graph}
\end{figure}
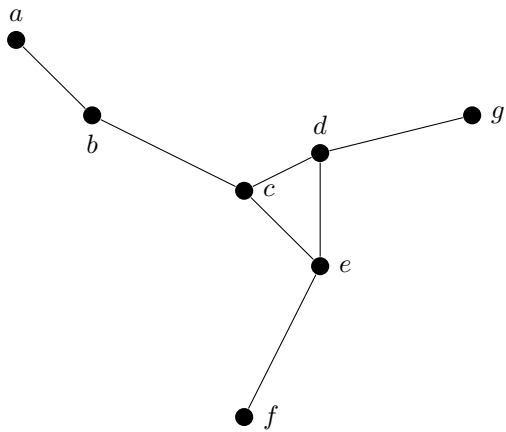

\subsubsection{Implication of those assumptions}

As described above the model implies several properties:
\begin{itemize}
    \item All combinations of traits are possible, thus we cannot have a trait whose existence depends on the other traits.
     \item All the traits behave similarly, which means mixing different types of data can lead to errors in the inference.
     \item It is possible for a trait to disappear in every \locus under our model. While in practice such a trait would not be recorded.
     \item A trait present in several \loci has necessarily been shared from single an initial point. It is not possible to have several sources for the same trait.
\end{itemize}

To account for the penultimate point, we will introduce the notion of \emph{registered} trait to designate traits that exist in at least one \locus. The last point is not probable for certain datasets, and \ourmethod would be inappropriate for them. For example, some cultural datasets have broad descriptions of traits (\textit{e.g.} ``agriculture is practised'') for which we cannot exclude several sources exist.

\subsection{Notations}

The parameters of the model are:

\begin{itemize}
    \item $\nu$ the rate at which the trait at a site disappear
    \item $\mu$ the rate at which a trait disappears on an edge
    \item $\mathfrak{G} = (E,V,d)$ the graph defined as the set of edges, vertices and the lengths of the edges. Note that if we define $d(e) = \infty$ for $e \notin E$, we remove the need to mention $E$.
\end{itemize}

We introduce notations for the latent variables:

\begin{itemize}
    \item $S_i$ is the source of trait $i$, we write $\mathbf{S}=(S_i)_i$
    \item $t_i$ the time at which $i$ appeared measured positively towards the past from $0$ the observation time.
    \item $\mathcal{P}_i \subset E$ a \emph{pattern}, the set of edges on which the trait $i$ \emph{can} propagate.
\end{itemize}

\subsubsection{Interpretation of the model}

An obvious interpretation, and our first interpretation, of what we chose to call \locus was merely a geographical place, the  ``distance'' matrix representing the physical difficulty to move from a place to another. But as the model is primarily designed for languages and cultural artifacts, we must underline that people, who speak languages and produce artifacts, can move. It is possible that a \locus corresponds to a certain population that moves in space. Conversely, we cannot either claim that a \locus is a single population, as human populations can merge and split. It is possible that two \loci represent a same population separated by some natural border that slowly grows apart by lack of exchange.

Similarly, the spread of a trait can be by mere exchange of ideas between populations, but also exchange of people themselves.

All of these questions are beyond the scope of the article, but we want to underline that this model should be use with care when trying to associate the results with other beliefs or information.

\subsubsection{A comparison with other interpretations of the wave model}

In \citep{Hartmann2023Wave}, the author propose an agent based generative model for the Wave model, where the geography of the area replaces the graph of \ourmethod. Learning a graph allows for more flexibility in the model, and one could argue that unless the data is collected continuously in space, learning a graph is actually easier.

\section{Parameters of interest}

The main parameter of interest is clearly $\mathfrak{G}$ as it impacts all the other parameters, and the uncertainty on $\mathfrak{G}$ have repercussions on the other parameters.

For example, the sources $\mathbf{S}$ are also interesting parameters, but if two \loci are extremely close in $\mathfrak{G}$, it is virtually impossible to identify precisely the source.

The evolution parameters $\mu$ and $\nu$ are more complex to identify, as it mostly depend on the age of the innovations studied, and further work is needed to assess their identifiability in practice.

Finally, $(t_i)$ will be integrated out, but can be inferred later conditionally on the other parameters. Independence of the innovations given the other parameter makes this parameter less interesting to learn directly.

\section{Computation of the Likelihood}

In this section we discuss theoretical and practical ways to compute the likelihood of the model defined above.

Our goal is to infer jointly $(\mu,\nu,\mathfrak{G},\mathbf{S})$. As $\mathfrak{G}=(E,V,d)$, $V$ is fixed and $d$ contains also the information of $E$, the likelihood writes:

\[ P( D_{i,\cdot}\mid \mu,\nu,d,\mathbf{S}) = \prod_{i=1}^mP( D_{i,\cdot}\mid \mu,\nu,d,S_i).\]

That is all the traits are independent given the latent parameters. We thus only need to compute:

\begin{equation}
    P( D_{i,\cdot}\mid \mu,\nu,d,S_i)  = \int P( D_{i,\cdot}\mid \mu,\nu,d,S_i,\mathcal{P}_i,t_i) f(t_i; \mathcal{P}_i,S_i,\mu,\nu,d) P(\mathcal{P}_i \mid d,\mu)\mathrm{d}t_id\mathcal{P}_i \label{eq:lkld1}
\end{equation} 

The question is now to compute $f(t_i; \mathcal{P}_i,S_i,\mu,\nu,d) \propto 1 \times P(R_i \mid t_i, \mathcal{P}_i,S_i,\mu,\nu,d) $ the intensity of the Poisson process that generates the \emph{registered} data. We write $\mathcal{R}_i$ the event that the trait $i$ produces a \emph{regisitered} data line. As we said previously, we assume the traits are born according to a  Poisson process with rate $1$, but it is not possible to register a trait absent of all the \loci. This second term $P(R_i \mid t_i, \mathcal{P}_i,S_i,\mu,\nu,d)$, corresponds to the probability that a trait born in $t_i$ produces a \emph{registered} line in the dataset. In particular, it is not possible to have $D(i,\cdot) = \mathbf{0}$. We could also imagine other registration conditions in future works.

To compute the quantity of Eq. \ref{eq:lkld1}, we first start by integrating over $t_i$.

\subsection{Integrating over $t_i$ given $\mathcal{P}_i$}

The death time of a trait on each branch follows an exponential distribution with pararameter $\mu$. So the probability that a trait propagates on the branch $e$ of length $d(e)$ is $1-\exp(-d(e) \mu)$.

A trait $i$ appears at $S_i$ at time $t_i$ following a Poisson process with constant rate. Note that under this assumption the number of trait is infinite, we will discuss that later. The trait will then spread along the edges. If we introduce $\mathcal{P}_i$ the pattern of the edges on which the trait \emph{can} propagate, the likelihood for one trait writes:

\begin{equation}
    P( D_{i,\cdot}\mid \mu,\nu,d,S_i,t_i,\mathcal{P}_i)= \prod_{j=0}^n \begin{cases}
        \e^{-\nu(t_i-d_{\mathcal{P}_i}(S_i,j)) } & \text{ if } D_{i,j}=1 \text{ and } d_{\mathcal{P}_i}(S_i,j)  \leq t_i;\\
        (1-\e^{-\nu(t_i-d_{\mathcal{P}_i}(S_i,j)) }) & \text{ if } D_{i,j}=0 \text{ and } d_{\mathcal{P}_i}(S_i,j)  \leq t_i;\\
        1  & \text{ if } D_{i,j}=0 \text{ and } d_{\mathcal{P}_i}(S_i,j)  \geq t_i; \\
        0 & \text{ if } D_{i,j}=1 \text{ and } d_{\mathcal{P}_i}(S_i,j)  \geq t_i.
    \end{cases}
    \label{eq:lkld_cond_tout}
\end{equation}

This expression is fairly simple but requires to compute distances on a graph, which is the main source of numerical complexity of \ourmethod. The fact that a trait can only spread once is represented by  $d_{\mathcal{P}_i}(S_i,j)$, as if a trait spread it only spreads through the shortest path (knowing the pattern).

When integrating over time, we need to integrate the quantity \ref{eq:lkld_cond_tout} against the Poisson process that generates the data at $S_i$. Although the apparition rate is constant equal to $1$, it is not possible to observe a line of data that would be only constituted of $0$s. Thus we introduce: $\mathcal{R}_i$ the event that the innovation $i$ is recorded, that is that there exists at least one \locus at which it exists when the data is recorded. In other words, it is not possible to observe $D_{i,\cdot} = \mathbf{0}$.




The probability of the trait to be recorded is then:

\[ P(\mathcal{R}_i \mid S_i,d,t,\mu,\nu) = 1 - \prod_{j,v\leq t} \exp(-\nu (t - d_{\mathcal{P}_i}(S_i,j)) \]

In other words, the process in that generates recorded traits in $S_i$ has the following intensity:

\begin{align*}
    f_{\mathcal{P}_i,d,\mu,\nu,d}(t) &=  P(\mathcal{R}_i \mid \mu,\nu,S_i,t,\mathcal{P}_i) \left( \int_0^\infty  P(\mathcal{R}_i \mid \mu,\nu,S_i,s,\mathcal{P}_i) \mathrm{d}s \right)^{-1} \\
    &:= P(\mathcal{R}_i \mid \mu,\nu,S_i,t,\mathcal{P}_i)/K_{\mu,\nu,\mathcal{P}_i,S_i}.
\end{align*}

And thus, the likelihood of a line of the dataset writes, integrating out the time in \ref{eq:lkld_cond_tout}:

\begin{align*}
    & \int_{0}^{\infty} f_{\mathcal{P}_i,d,\mu,S_i}(t_i) P( D_{i,\cdot} \mid \theta,d,S_i,t_i,\mathcal{P},\mathcal{R}_i)\mathrm{d}t_i \\
    &= \frac{1}{K_{\mu,\nu,\mathcal{P}_i,S_i,d}}\int_0^\infty P(\mathcal{R}_i \mid \mu,\nu,d,S_i,t_i,\mathcal{P}_i,) P( D_{i,\cdot} \mid \mu,\nu,d,S_i,t_i,\mathcal{P}_i,\mathcal{R}_i)\mathrm{d}t_i \\
    &= \frac{1}{K_{\mu,\nu,\mathcal{P}_i,S_i,d}}\int_0^\infty P( D_{\cdot, i}, \mathcal{R}_i\mid \mu,\nu,d,S_i,t_i,\mathcal{P}_i)\mathrm{d}t_i \\
    &= \frac{1}{K_{\mu,\nu,\mathcal{P}_i,S_i,d}}\int_0^\infty P( D_{i,\cdot}\mid \nu,d,S_i,t_i,\mathcal{P}_i)\mathrm{d}t_i,
\end{align*}

where we removed the dependency in $\mu$, as the data only depends on $\mu$ through $\mathcal{P}_i$.

\subsection{Practical computation of the likelihood}

\label{sec:integraletemps}

The previous term require to integrate over all possible apparition times. Note that if we define $(s_0,s_1,\dots,s_n)$ as the incrasingly sorted vector of the $(d_{\mathcal{P}_i,S_i}(j))_{j \in V}$, so that $s_0 = d_{\mathcal{P}_i,S_i}(S_i) = 0$, we can write the integrals:

\[K_{\theta,\mathcal{P},S_i,d} = \int_{0}^{\infty} P(\mathcal{R}_i \mid \nu,S_i,s,\mathcal{P}_i) \mathrm{d}s, \]

and

\[  \int_0^\infty P( D_i\mid \mu,\nu,d,S_i,t_i,\mathcal{P})\mathrm{d}t_i, \]

as

\begin{align}
    \int_{0}^{\infty} P(\mathcal{R}_i \mid S_i,d,\mathcal{P}_i,s) \mathrm{d}s &= \int_{s_n}^{\infty}P(\mathcal{R}_i \mid S_i,d,\mathcal{P}_i,s) \mathrm{d}s + \int_{s_{n-1}}^{s_n} P(\mathcal{R}_i \mid S_i,d,\mathcal{P}_i,s) \mathrm{d}s + \dots  \nonumber\\
    & + \int_{s_1}^{s_2} P(\mathcal{R}_i \mid S_i,d,\mathcal{P}_i,s) \mathrm{d}s+ \int_{0=s_0}^{s_1} P(\mathcal{R}_i \mid S_i,d,\mathcal{P}_i,s) \mathrm{d}s, \label{eq:lklcomput}
\end{align}

and

\begin{align}
    \int_{0}^{\infty} P( D_i\mid \mu,\nu,d,S_i,t_i,\mathcal{P})\mathrm{d}t_i &= \int_{s_n}^{\infty}P( D_i\mid \mu,\nu,d,S_i,t_i,\mathcal{P})\mathrm{d}t_i + \int_{s_{n-1}}^{s_n}P( D_i\mid \mu,\nu,d,S_i,t_i,\mathcal{P})\mathrm{d}t_i + \dots  \nonumber\\
    & + \int_{s_1}^{s_2} P( D_i\mid \mu,\nu,d,S_i,t_i,\mathcal{P})\mathrm{d}t_i+ \int_{0=s_0}^{s_1} P( D_i\mid \mu,\nu,d,S_i,t_i,\mathcal{P})\mathrm{d}t_i.\label{eq:lklcomput2}
\end{align}

Then, as both these integrands are polynomial in $y = \e^{-\mu t_i}$:

\begin{align*}
    \int_{s_r}^{s_{r+1}} P(\mathcal{R}_i \mid \nu,S_i,d,\mathcal{P}_i,t_i) \mathrm{d}t_i &= \int_{s_r}^{s_{r+1}} 1 - \prod_{j=0}^r(1- \e^{\nu s_j}\e^{-t_i\nu}) \mathrm{d}t_i  \\
    &= \int_{\e^{-\nu s_r}}^{ \e^{- \nu s_{r+1}}} \frac{1 - \prod_{j=0}^r(1- \e^{\nu s_j}y)}{-\nu y} \mathrm{d}y = \int_{\e^{-\nu d_j}}^{ \e^{- \nu d_{j+1}}} \frac{1 - R_j(y)}{-\nu y} \mathrm{d}y \\
\end{align*} 

similarly

\begin{align*}
    \int_{s_r}^{s_{r+1}} P( D_i\mid \nu,d,S_i,t_i,\mathcal{P}_i)\mathrm{d}t_i&= \int_{\e^{-\nu s_r}}^{ \e^{- \mu s_{r+1}}} \frac{\prod_{j=0}^{r} \mathbf{1}_{D_{i,j}=1}\e^{\nu s_j}y +\mathbf{1}_{D_{i,j}=0}(1-\e^{\nu s_r}y)}{-\nu y} \mathbf{1}_{\forall \ell \geq r+1, D_{i,\ell}=0} dy \\
    &=\int_{\e^{-\mu d_j}}^{ \e^{- \mu d_{j+1}}} \frac{Q_j(y)}{-\nu y} \mathbf{1}_{\forall \ell \geq r+1, D_{i,\ell}=0} dy,
\end{align*}

These last polynomials can be computed recursively in order to find exact integral expressions:

\begin{align*}
    R_0 &= (1 - \e^{\nu 0}X) = 1 - X \\
    R_{r+1}&= R_r (1-\e^{\nu s_{r+1}}X). \\
    Q_0 &= e^{\nu 0}X = X \\
    Q_{r+1}&= Q_j(\mathbf{1}_{D_{i,j}=1}\e^{\nu s_{r+1}}X +\mathbf{1}_{D_{i,j}=0}(1-\e^{\nu s_{r+1}}X))
\end{align*}

Which in turns allow for simple implementation of the integral computation with exact formulae, keeping in mind that $\forall t \leq \min_{j, D_{i,j}=1}s_j,P(D_{i,\cdot}\mid \nu,d,S_i,\mathcal{P})=0$ as the data is impossible (the trait wouldn't have had enough time to reach all the sites at which it was observed).

\subsection{Unbiased estimation of the integral in $\mathcal{P}_i$}

To estimate the integral in \ref{eq:lkld1} over $\mathcal{P}_i$, we rely on a simple Monte-Carlo estimator:

\begin{align*}
    &P( D_{i,\cdot}\mid \mu,\nu,d,S_i)  \\
    \simeq &P(\mathcal{P}\neq E \mid \nu,d)^{-1}\sum_{k=1}^{K} P( D_{i,\cdot}\mid \mu,\nu,d,S_i,\mathcal{P}_k) + P(E \mid \nu,d)^{-1}P( D_{i,\cdot}\mid \mu,\nu,d,S_i,E)
\end{align*}
where $(\mathcal{P}_k)_k$, are simulated from a density proportional to

\[    \exp\left( \sum_{e \in \mathcal{P}} -\mu d(e) \right)\mathbf{1}_{\mathcal{P}\neq E}.\]

This estimation is unbiased, as the computation from \ref{sec:integraletemps} is exact up to numerical error. We always add in the sum the case $\mathcal{P}=E$ so that the likelihood estimate is never null.

\section{Prior distributions}

We chose on $\mu$ and $\nu$ exponential priors, as we want to favour low values of these parameters.

On the sources, we chose a uniform prior on all the \loci.

On the graph, we chose a prior on the form $\pi(\mathfrak{G}) \propto \exp(-\theta \vert E \vert) $, that is a prior that favours sparse graphs. We chose a uniform on $\mathbb{R}$ prior on the edge length. This choice is discussed in a supplementary material, and in practice did not lead to dramatically different results, despite the prior here being improper.


\section{Identifiability of the model}

Although the likelihood is formally identifiable if we fix  the rate at which new innovations appear (which we do from the start), practical identifiability is more complex to judge.

In particular, if three \loci, $A$, $B$ and $C$, are connected by short edges, the observations at those nodes can be very similar (depending on $\nu$ and the time of arrival of the innovation). The likelihood of the four models $A \leftrightarrow B \leftrightarrow C$, $A \leftrightarrow C \leftrightarrow B$, $C \leftrightarrow A \leftrightarrow B$ $A \leftrightarrow B \leftrightarrow C \leftrightarrow A$ will be almost the same.

This extends to larger groups of \loci, which can lead to more difficult intrepretation of the posterior distributions and at least more care in the analysis of the outputs: two seemingly different outputs can merely be two almost equivalent graphs. This problem is further exacerbated by the mixing properties of the numerical methods.

\section{Numerical methods}

We implemented, a Metropolis-Rosenbluth-Rosenbluth-Teller-Teller-Hastings-within-Gibbs algorithm. We provide in the supplementary material a more precise description of the Gibbs moves proposed for each of the parameters. Broadly speaking, we spend most of the time updating the topology of the graph, we update exactly the position of the sources by computing the full posterior distribution over the source for each data point, and we propose joint movements on the sources and the topology of the graph.

More details on how each of the moves is implemented can be found in supplementary material.

We use parallel implementation in the computation of the likelihood, as given the parameters each data point is independent, we could also parallelise the computation of the sum over the sampled patterns but did not appear to need it.

\subsection{Computational cost of \ourmethod}

The cost for estimating the state likelihood can be written, in terms of number of operations as:

\[ \mathcal{O}(m \times n \log(n) \times K) \]

that is more than linear in the number of languages and linear in the other parameters. This term is explained by the requirement for us to compute all the distances to the source and order them, and then to compute the polynomials and integrating them.

The size of the parameter space is quadratic in $n$ and linear in $m$.

Overall, we recommend keeping the number of studied \emph{loci} under 20, unless working with considerably large parallel architectures.

\section{Numerical experiments}

We present here results on several datasets, additional results are presented in the supplementary material.

We ran all the simulations on 41 Intel(R) Xeon(R) Gold 6338 CPU @ 2.00GHz.

\subsubsection{Remark: graphical representation of the posterior distribution}

We chose to represent the posterior on $\mathfrak{G}$ as a posterior distribution on the matrices. Although this is the simplest representation and the most straightforward, we do not believe it to be an accurate representation of the posterior information. We nevertheless chose to keep this representation to avoid adding the introduction of another representation that would depend on additional parameters.

The main feature that is harder to detect on the current representation of the posterior is the interaction of the nodes. For example, if nodes $A-D$ are all extremely close and $E$ is related to any of those, the posterior when $E$ is connected to any of those node will be of similar value, thus diluting the posterior distribution.

This question and more generally the accurate representation of similar objects should be investigated in the future.

\subsection{Simulated datasets}

\begin{figure}
    \centering
\includegraphics[width=0.9\textwidth]{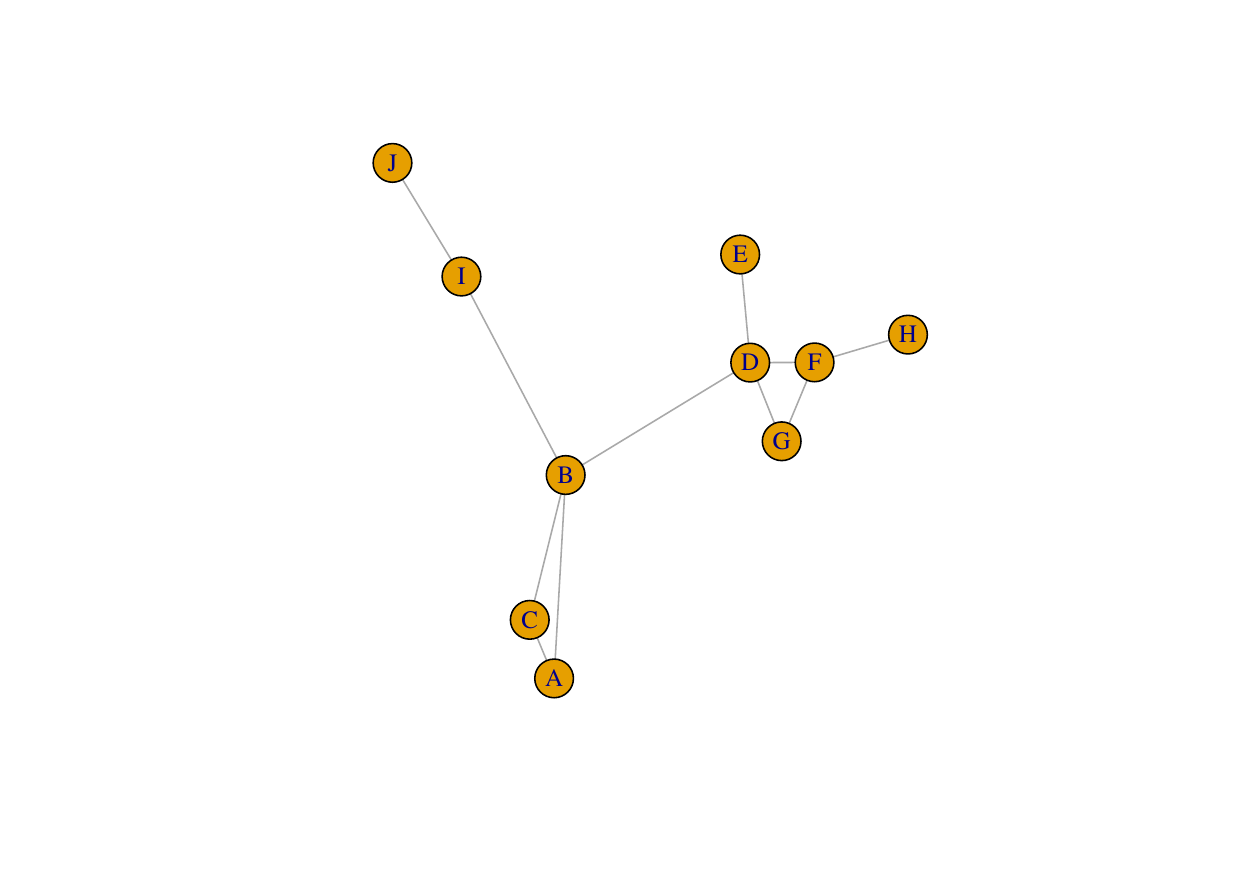}
\includegraphics[width=0.7\textwidth]{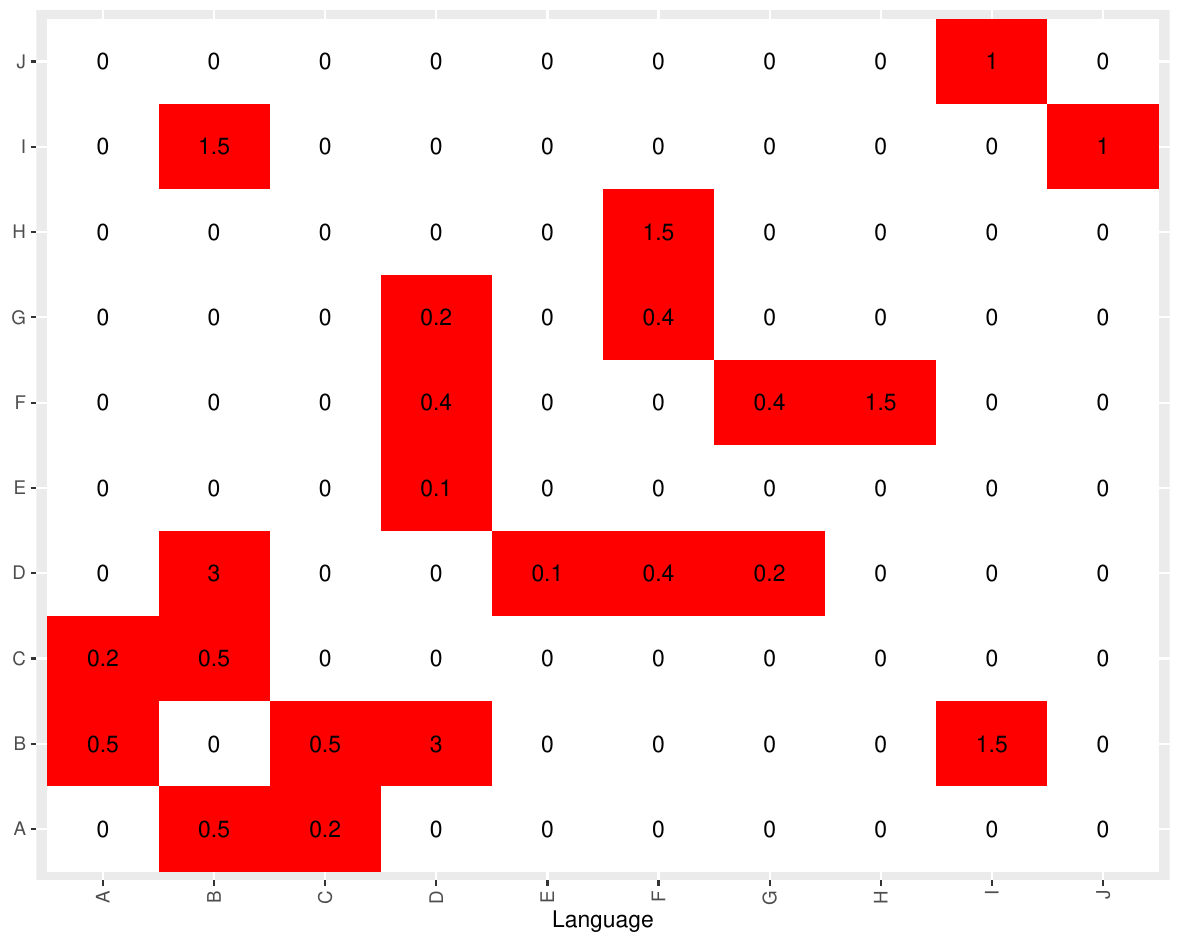}
\caption{True graphed of the simulated results: graphical representation and adjacency matrix.}
    \label{fig:simu_graph}
\end{figure}

First, in order to test the model and method, we propose to run inference on a simulated dataset. We simulated 200 recorded traits from the graph represented in Figure \ref{fig:simu_graph}, with parameters $\mu=0.3$, $\nu=0.01$ and traits appearing uniformly between $0$ and $3.5$. We used an exponential prior centered around the true value for $\nu$ and $\mu$, and on the graph the prior above with parameter $\theta=0.5$.

We ran \ourmethod for 40000 MCMC steps, which lasted around half a day. To check convergence of the chain, we plotted the (subsampled) likelihood along the chain. It is clear that in some cases the chain remains stuck in local maximum, the replicas of MCMC, which are easy to spot, are not included in the results, and account for around a fifth of the simulations. This can be explained by the difficulty to move from one state to another given our choices of moves, but overall this happened seldom in our experiments. 

\begin{figure}
    \centering
    \includegraphics[width=0.7\linewidth]{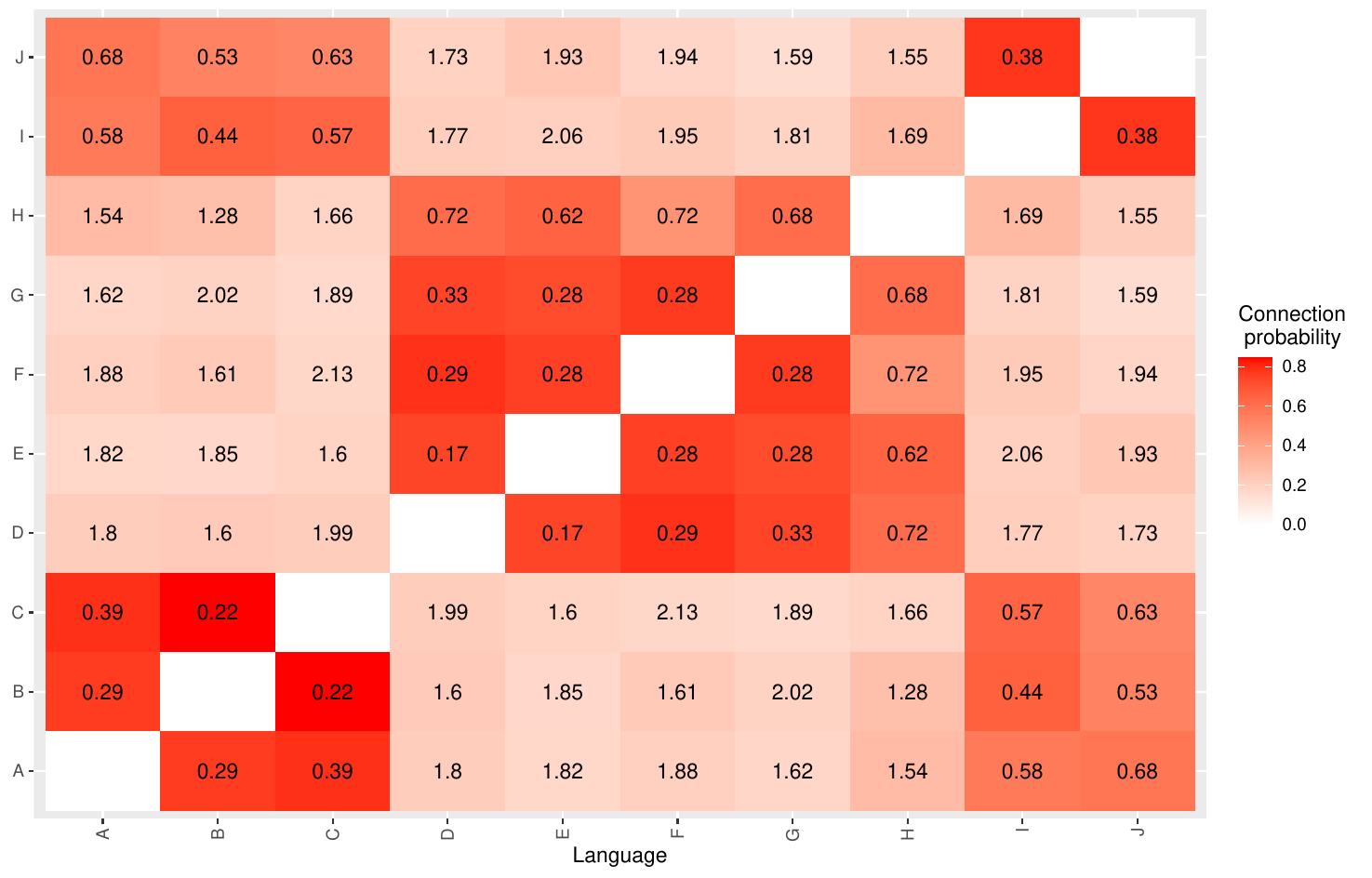}
    \includegraphics[width=0.7\linewidth]{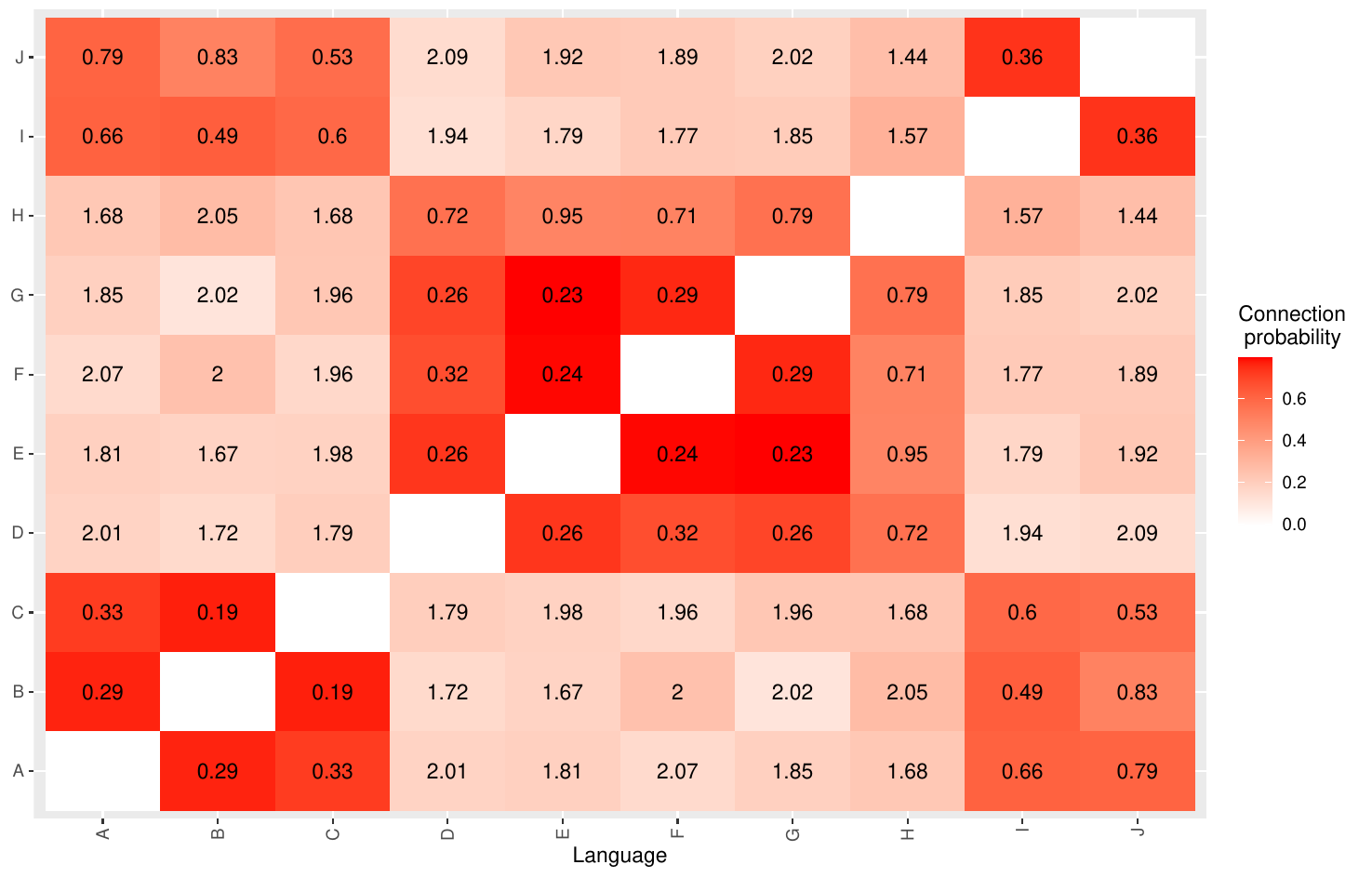}
    \caption{Posterior distribution on the graph for the simulated dataset. Two independent realisation of \ourmethod.}
    \label{fig:simulated}
\end{figure}

The results, presented in Figure \ref{fig:simulated}, are globally positive. In particular we observe the three groups of vertices, and adequate topology. Of course it is not possible to identify exactly the arrival point of a long edge onto a tightly connected group, which explain the ``block'' of \loci D to H. The uncertainty, that is most likely due to the uncertainty of the model, as it is observed in all the different replicas with different priors (see supplementary material, explains the results: A, B and C are all connected, which in turn means that J and I are connected with any of A, B and C, but with uncertainty as for which is connected to which. The same can be said about the D to H group. Overall, it seems \ourmethod can only reconstruct blocks of closely related \loci.

Notably, the long D-B connection is not reconstructed, which is easily explained by the length of the edge. It is actually more parcimonious to explain the connection between the two groups though the I-J group. It is anyway relatively unlikely an innovation would require this edge to be transmitted.


\subsubsection{Convergence and stability of the method}

\begin{figure}
    \centering
    \includegraphics[width=0.5\linewidth]{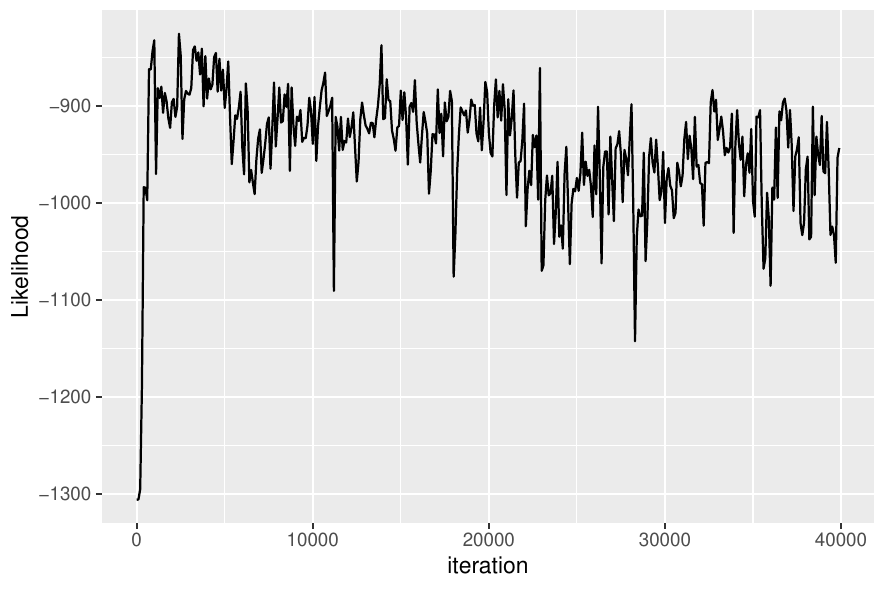}
    \caption{Plot of the likelihood along the chain of a simulation of \ourmethod on the simlated dataset}
    \label{fig:simu-lkld}
\end{figure}

Numerically, the method converged fairly quickly, as can be seen from the plots of the likelihood (Fig \ref{fig:simu-lkld}). In some cases and other examples, the chain seems to get stuck in lower likelihood state, temporarily or indefinitely. This is easy to spot on the simulation, and only occured seldom.

We can compute an ESS for each of the values of the connection matrix. These ESS tends to be low for the low-connection-probability edges (around a hundred) and higher for the other (several hundreds). Each of the moves only impacting one or two edges, these numbers seems decent.

\subsubsection{Parameter reconstruction} The  parameter to estimate outside of the graph are $\mu$ and $\nu$, the rate of disappearing of the traits. In our experiment, $\mu$ is slightly overestimated, while $\nu$ is slightly underestimated. Overall the moves to update the parameters were seldom proposed and accepted. The relative inefficiency of the method in recovering these parameters do not appear to impede the ability of \ourmethod to recover the main parameter of interest: the graph.

\subsubsection{Sources of the traits} these are obtained as additional output, we can compare the inferred sources to the true ones. In our simulations, only a little more than $55\%$ of the 0.5 confidence intervals for each source contained the true source. Nevertheless, if we define the three groups observed in the graph (A-C ; D-H; I-J), \ourmethod recovers the position of the source in the right group in more than 80\% of the cases.

It seems normal, given the shortness of the edges in each of the group that the source cannot be detected with certainty. In practice, I the topology of the graph first needs to be studied so that inference on the origin of the innovations can be presented in a meaningful way.

\subsection{Real dataset: Vanuatu languages}

\label{sec:resvanuatu}

\begin{figure}
\vspace{-2.5cm}
\hspace*{-\dimexpr\oddsidemargin+1in\relax}\makebox[\paperwidth]{
    \centering
    \includegraphics[width=1.2\linewidth]{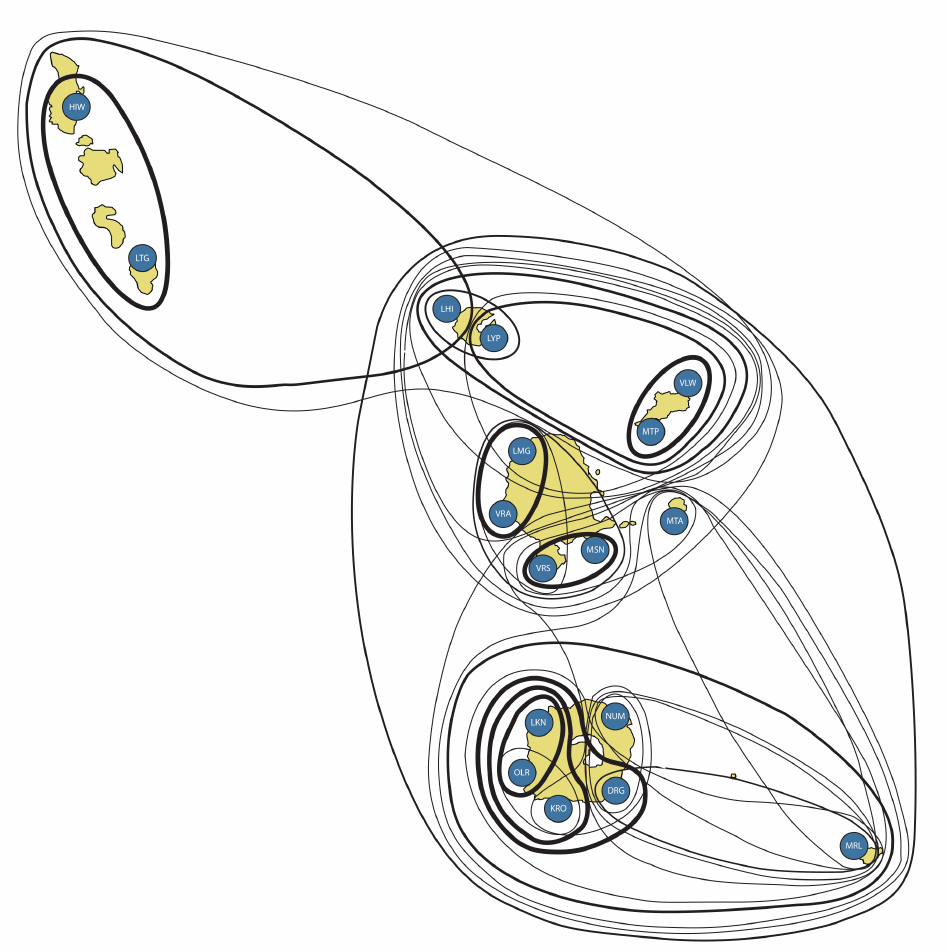}}
    
    \hspace*{-\dimexpr\oddsidemargin+1in\relax}\makebox[\paperwidth]{
    \includegraphics[width=1.2\linewidth]{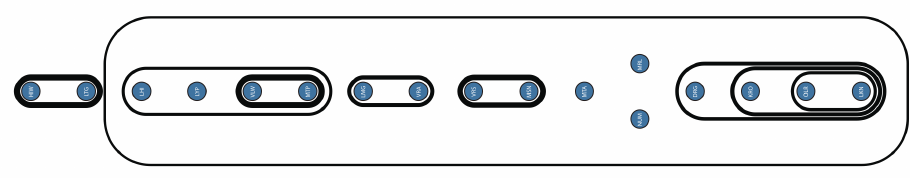}}
    \caption{Results from \citet{kalyan2019understanding} on the North Vanuatu languages on a geographical map (top), and summarised output with weakest isoglosses removed (bottom). The language names read as follows: {\sc hiw} Hiw, {\sc ltg} Lo Toga, {\sc lhi} Lehali, {\sc lyp} Löyöp,
{\sc vlw} Volow, {\sc mtp} Mwotlap, {\sc lmg} Lemerig, {\sc vra} Vera’a, {\sc vrs} Vurës, {\sc msn} Mwesen, {\sc mta} Mota, {\sc num}
Nume, {\sc drg} Dorig, {\sc kro} Koro, {\sc olr} Olrat, {\sc lkn} Lakon, {\sc mrl} Mwerlap.}
    \label{fig:resultatsKalyan}
\end{figure}

The dataset presented in \citet{kalyan2018freeing} (although made available by personal communication) contained lexical, morphological, syntaxic and phonological traits in north Vanuatu languages. In \citet{kalyan2019understanding} and \citet{kalyan2018freeing} the authors have proposed a descriptive representation of the traits to infer isoglosses, that is lines that split the different languages depending on the presence or absence of some traits. These isoglosses are often assumed to derive from the waves. We reproduce in Figure \ref{fig:resultatsKalyan} the results from the authors, we can observe that the overall structure of the data does not exclude that a Tree model be more adapted to explain the current observations. We nevertheless expect, given the geographical features of the area, that the languages of the central islands be the connexion between the more distant languages.

The dataset constituted of a mixture of lexical, phonological, grammatical and syntaxic traits. We only used in this example lexical traits, as the other were less numerous and we prefer to not use the same model for different types of traits.

We ran \ourmethod for 40000 steps with the same settings and priors as in the previous example. Given the number of traits and languages, the computation took a full day. We compared the results over several replicas, with a few failing to converge in the computational time (as seen on a plot of the likelihood), and got similar results. This can be easily detected by checking the convergence of the method.

\begin{figure}
    \centering

    \includegraphics[width=\linewidth]{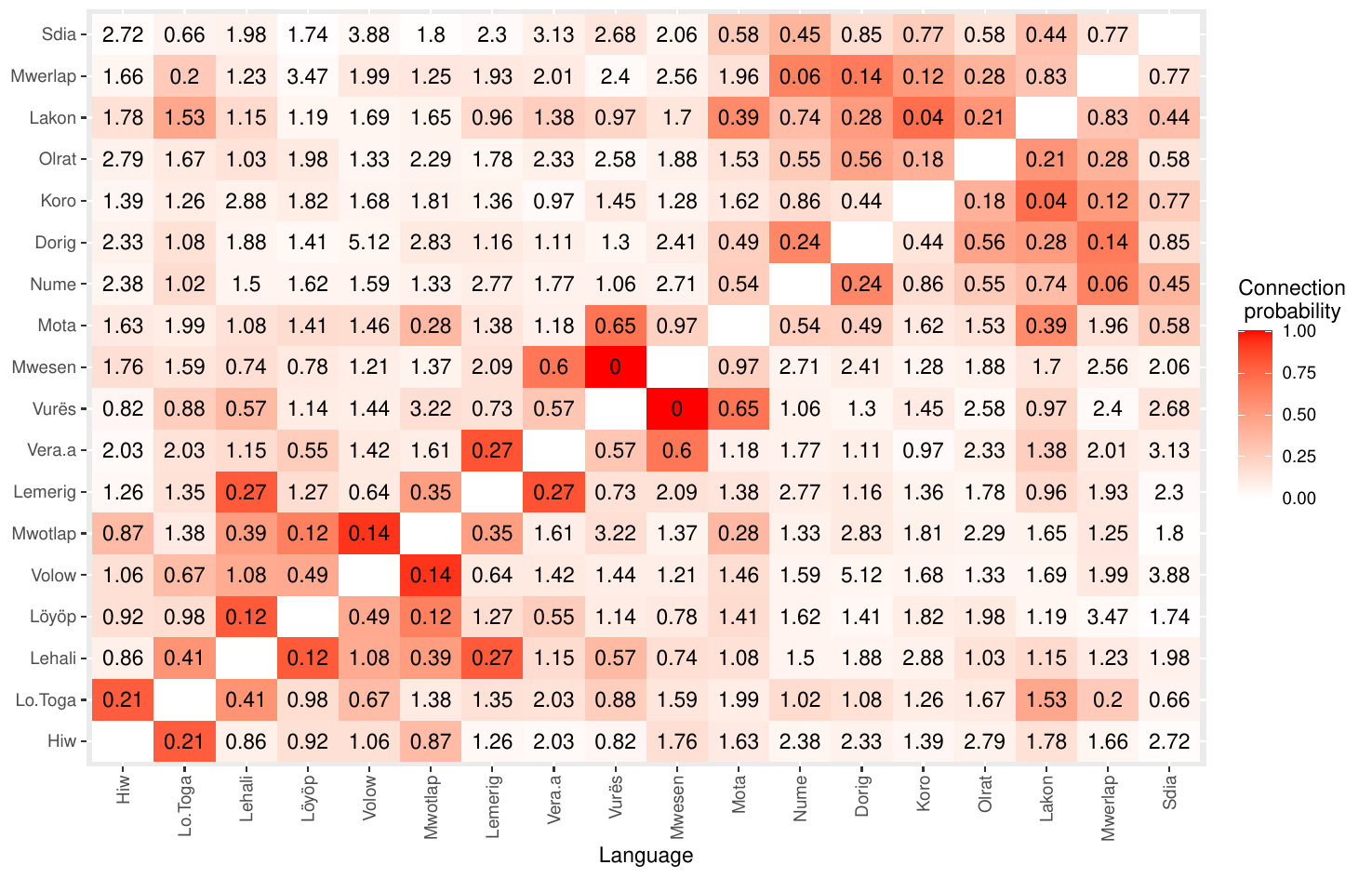}
    \includegraphics[width=\linewidth]{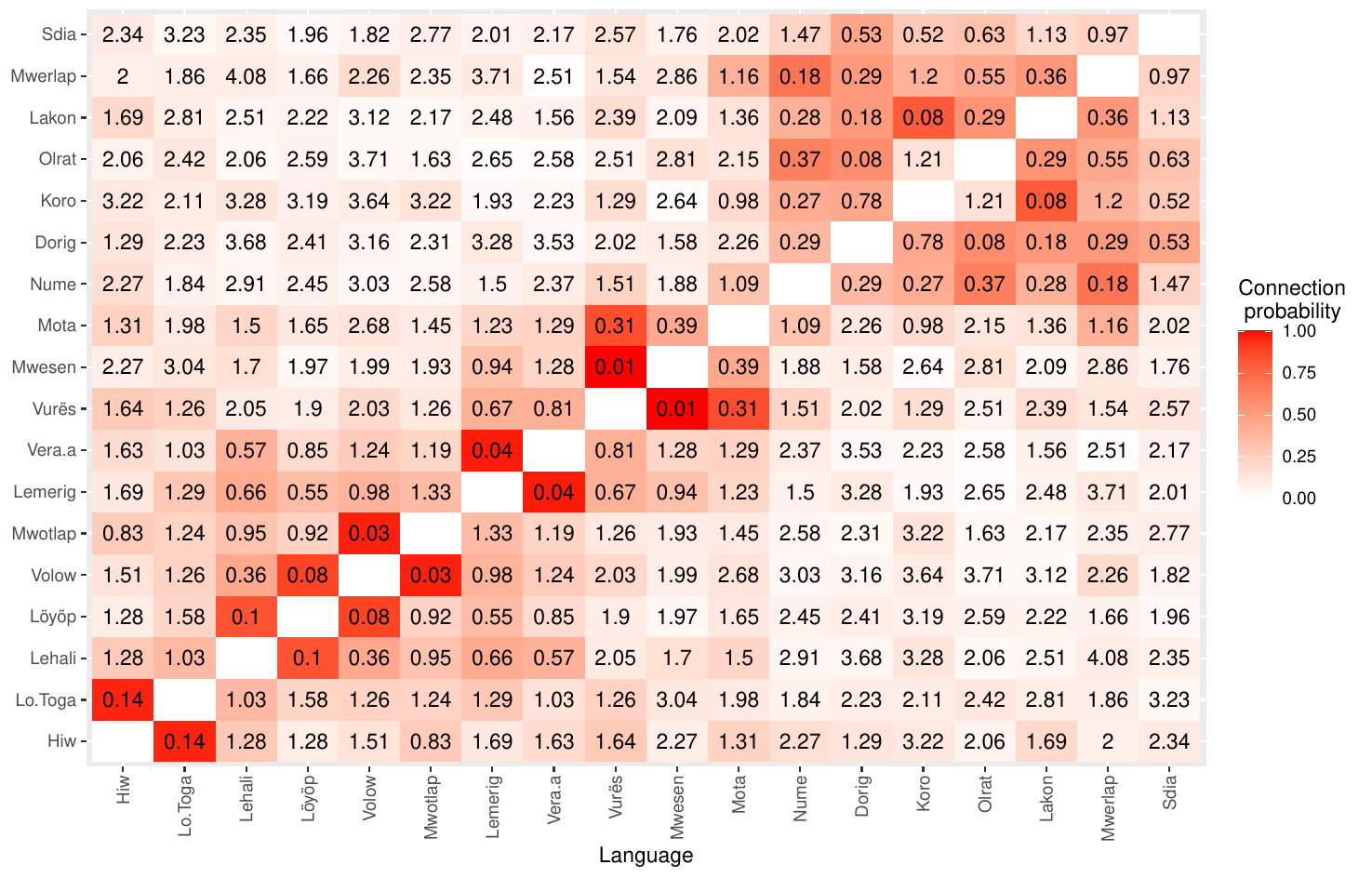}
    \caption{Results from \ourmethod on the North Vanuatu dataset.}
    \label{fig:resultsvanuatu20k}
\end{figure}

\begin{figure}
    \centering

    \includegraphics[width=\linewidth]{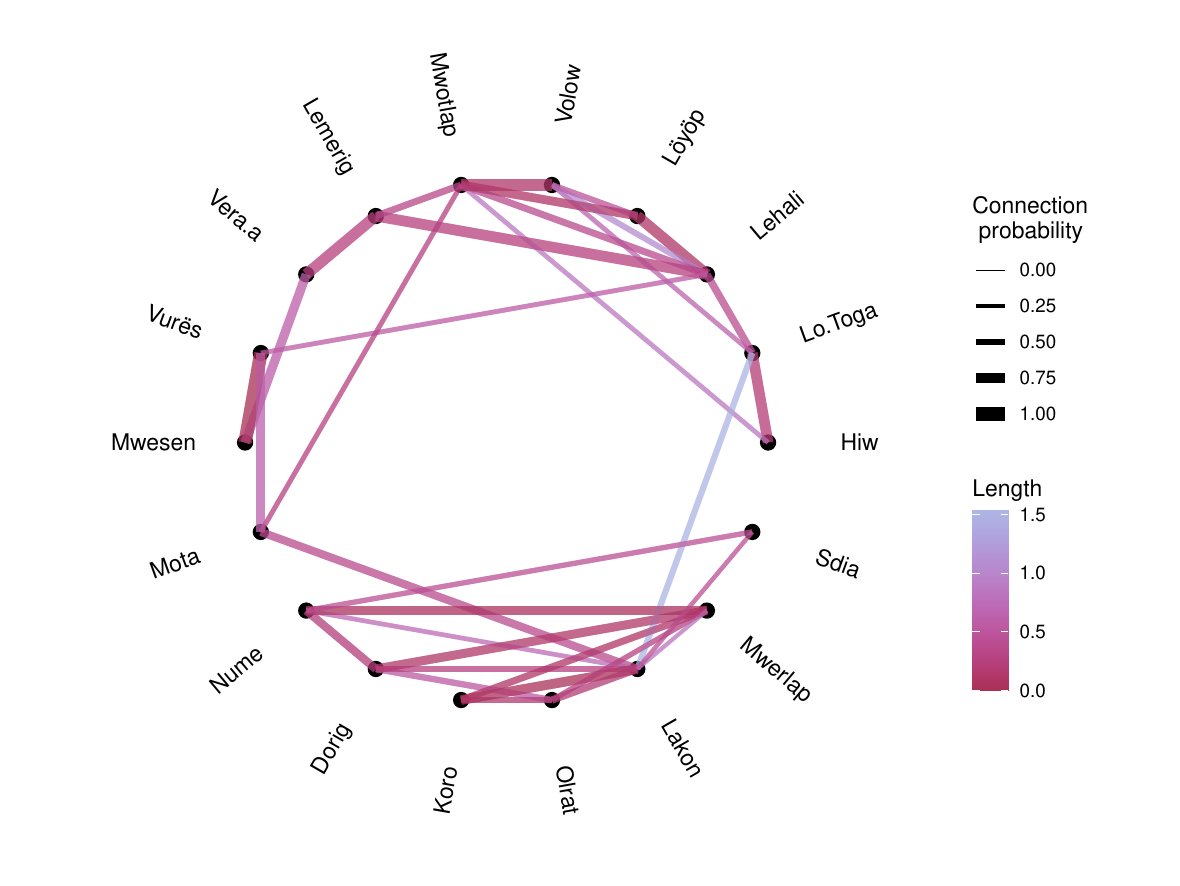}
    \includegraphics[width=\linewidth]{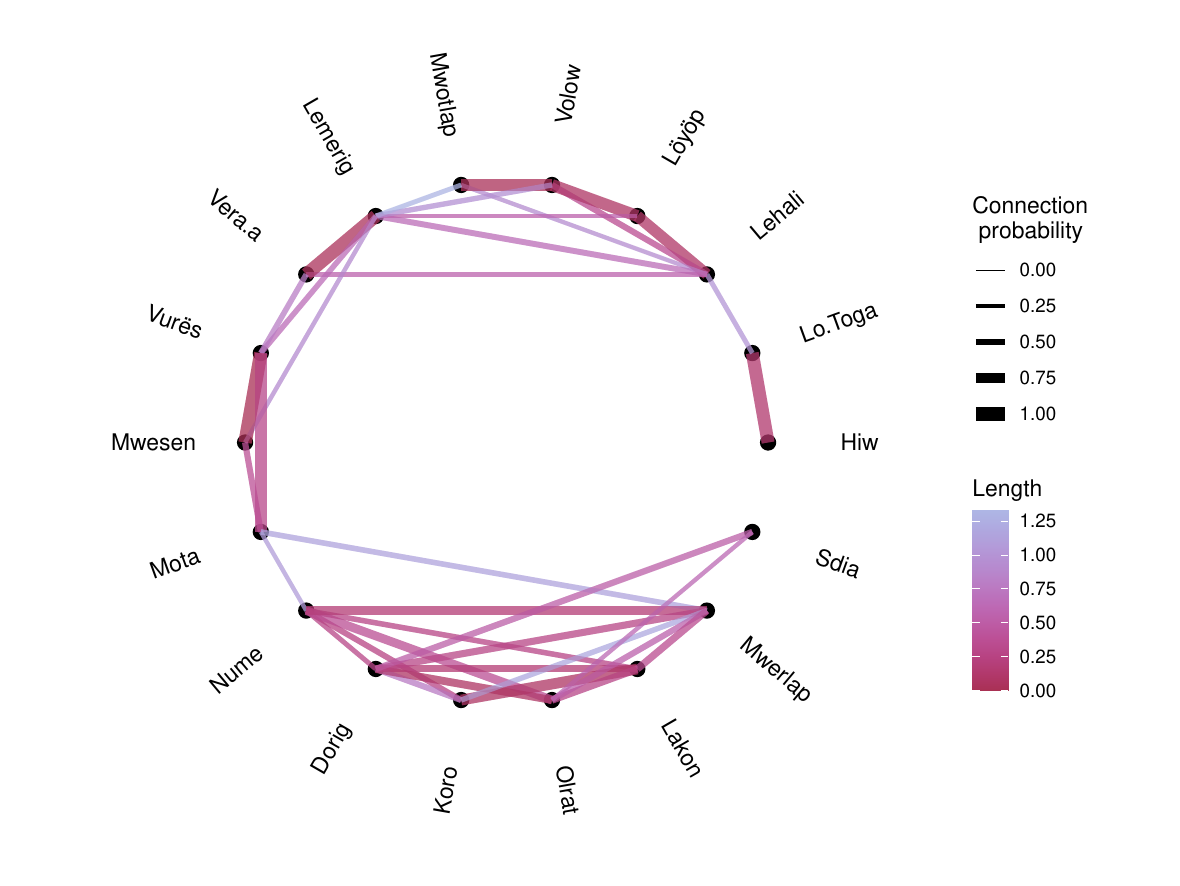}
    \caption{Results from \ourmethod on the North Vanuatu dataset with edges with posterior probability lower than $0.3$ removed.}
    \label{fig:resultsvanuatu20knetwork}
\end{figure}

The results, presented in \ref{fig:resultsvanuatu20k} and \ref{fig:resultsvanuatu20knetwork} are similar to the descriptive reconstruction from \citet{kalyan2019understanding}. We can see the central role of some of the languages as crossroads, for example Mota (MTA) which connects  the central group (with Vurës/VRS) the northern group (Lo Toga/LTG) and the southern group (Mwerlap/MRL), although the point of connection is more uncertain. The uncertainty on the results are always concerning the connections of the groups. Similarly, Lehali/LHI although closer to the north group seems to be a connection point with the more central group.

Overall, the graphical representation shows that the general gradient of innovations from an end of the archipelago to the other, that was hypothesised by specialists and found in the previous studies is confirmed by our simulations.

\subsection{Real dataset: Greek Dialects}

\begin{figure}
    \centering\includegraphics[width=\linewidth]{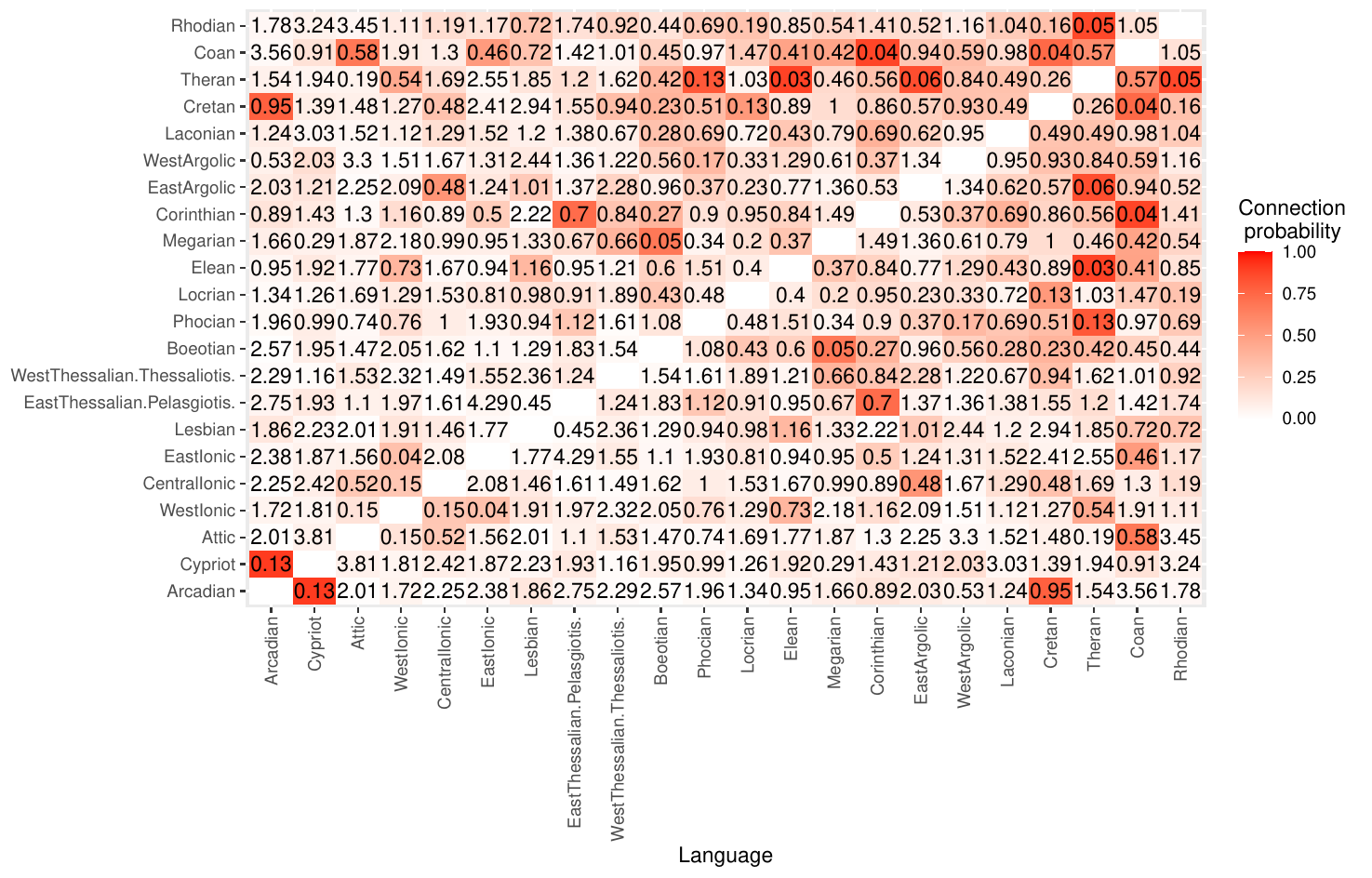}
    \includegraphics[width=\linewidth]{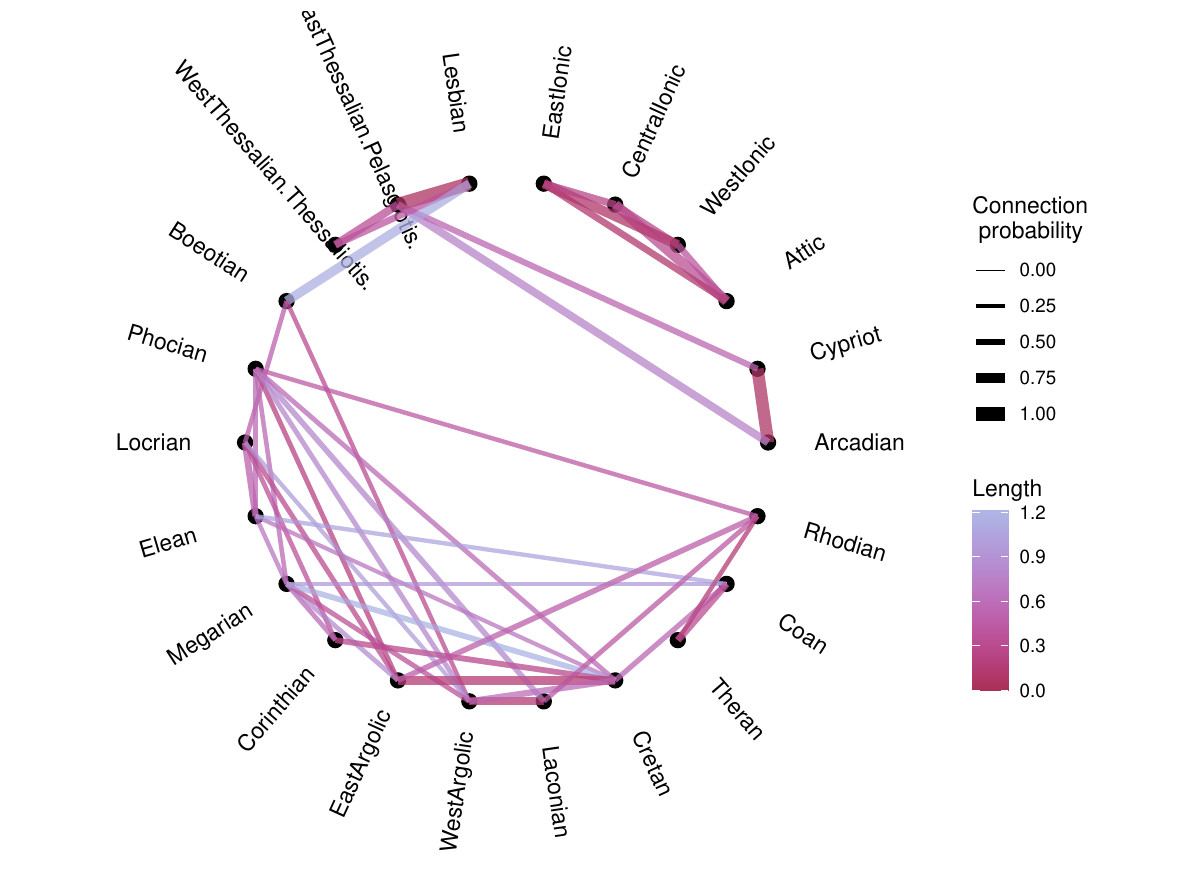}
    \caption{Results on the Greek dialects dataset. Top posterior on the individual edges, bottom graph representation with low probability edges removed}
    \label{fig:grec}
\end{figure}

In \citet{SKELTON2015Borrowing}, the author publishes a complete data of morphological and lexical innovations in Greek dialects. After some adaptation of the dataset with advice from the author, we ran the method with the same (hyper)-parameters as other experiments. The results are constant among replicas and are presented in Figure \ref{fig:grec}. Although there seems to be high uncertainty on the output, we can observe three groups that corresponds to the most frequently defined groups of dialects: The Ionic group (Attic and Ionian), the Arcadio Cypriot, and the Lesbian-Thessalonian group. Some other pairs are quite close, such as the Cretan and Argolic dialects.

\subsection{Real dataset: Loom techniques in the Kra-Dai people}

\begin{figure}
    \centering
    \includegraphics[width=\linewidth]{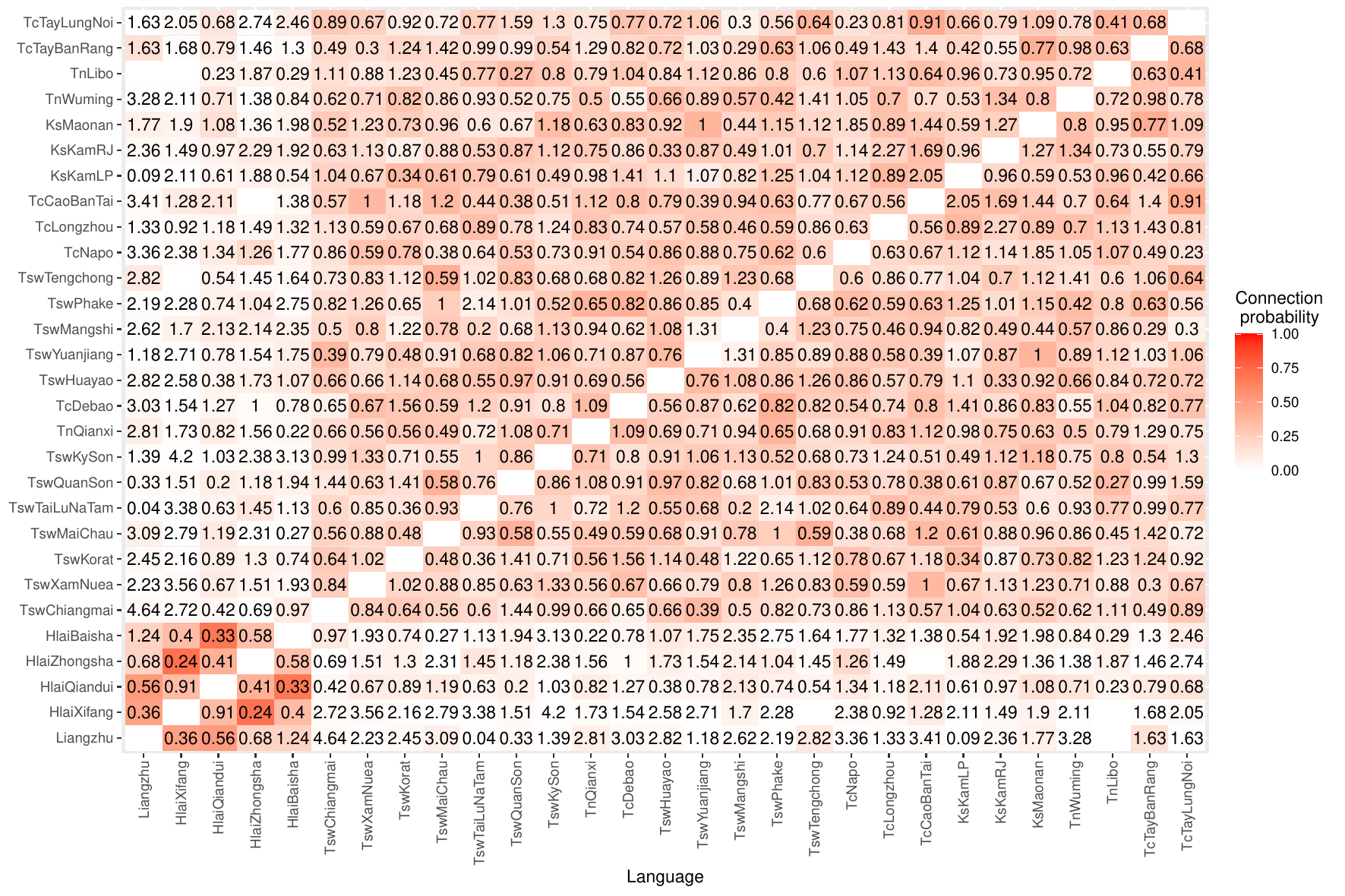}
    
    \includegraphics[width=\linewidth]{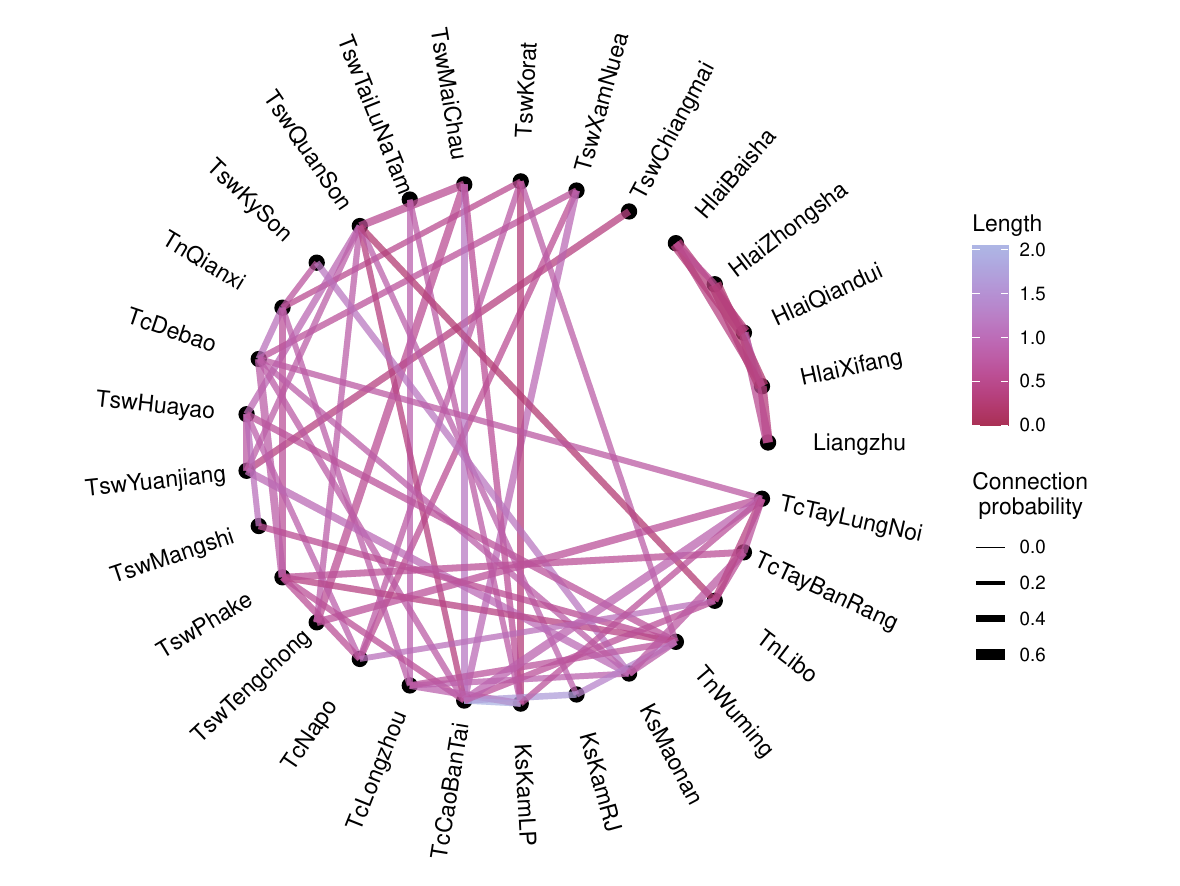}
    \caption{Results on the Kra-Dai looms dataset.}
    \label{fig:Looms}
\end{figure}

In \cite{buckley2025contrasting}, the author compare evolution of languages and Loom techniques on two trees and then together. Here, we chose to run \ourmethod on the loom dataset, that was binarized by the original authors. The tree on this dataset was relatively loosely defined, with low probabilities for each of the split in the consensus tree except at one point. The results presented in \ref{fig:Looms} show indeed this uncertainty: except a small outgroup clearly detected, no clear posterior links between the groups appears.

Overall, although the only group detected is known from other

\section{Discussion and future works}

This model a first attempt to formalize the Wave model as conceptualised in linguistics, there are still many problems to solve and many directions to explore. The results, without much tuning of the parameters, are nevertheless close to what is expected from specialists knowledge. We expect further refinement of the model and more careful choice of the data will allow to produce even more significant results. In terms of future directions we can list:

\paragraph*{Posterior representation} The current posterior representation is not satisfying on several aspects, but we were unable to propose something more convincing and as simple. Among the possible directions, either works similar to \citet{balocchi2025understanding} or something \textit{ad hoc} would be an improvement.

\paragraph*{Minor improvements} 
We could differentiate in the dataset and in some cases two types of $0$'s, first a trait that has never been adopted, and secondly a trait that has disappeared in the present. An example of such a trait would be a word of vocabulary in a language that died out in every-day language (e.g. \emph{choir} in french) but still exist in other words (\emph{chûte, cascade}), so that we know for sure that this trait \emph{was} present. Another simple change would be to remove the symmetry assumption, either in the length of the edges or overall. This would allow to represent that some nodes are more influential than others, in terms of languages, we can think about the influence of a culturally dominant language (e.g. classical chinese).

\paragraph*{Numerical improvements} \ourmethod suffers from large numerical complexity and computational cost. Although several improvements were made on this side, we can foresee several future changes, mostly related to the computation of the likelihood. Proposing more reliable methods and better mixing proposals would also be beneficial, for example using SMC-like methods. This would also allow to compute marginal likelihoods.

\paragraph*{Changes in the model} Simple improvements of the model would be to allow for more possible uncertainties, for example to allow for different types of data to spread on the same graph. Another possibility, which would require to rework the entire model, at the point of being an almost independent work, would be that the distance between \loci would depend on the number of common innovations. This model would allow us to describe longer term evolutions, and thus might be more adapted to indo-european datasets than dialects. 

\paragraph*{Applications} This model can be applied in several contexts. Although linguistical data is particularly adapted, we explored the possibility of working on cultural or archaeological datasets. The main issues is the requirement of observing \emph{fossilised} \loci and non-binary innovations (\emph{e.g.} different degrees of a same trait).

\section*{Acknowledgements}

This work has made use of the resources provided by the Edinburgh Compute and Data Facility (ECDF).

We would like to thank Robin Ryder, Cecilia Balocchi and Antoine Diez for their advice and proofreading, Christina Skelton for discussions on the Greek Dialects dataset, Siva Kalyan and Alexandre François for discussions and the north Vanuatu dataset, and Václav Hrnčíř for discussions on cultural datasets.

\bibliographystyle{plainnat} 
\bibliography{mabiblio}  

\newpage

\appendix

\section{Additional details on the numerical methods}

\label{sec:supp:nummeth}

As the method used is a fairly simple MCMC-within-Gibbs algorithm, the only parameter we can truly tune is the choice of the Gibbs move. Currently we used four types of moves:

\begin{itemize}
    \item Moves on the topology of the graph (adding, removing and moving edges);
    \item Joint moves on the topology and the sources (same as above);
    \item Moves on the sources;
    \item Moves on the parameters.
\end{itemize}

The joint moves were by far the more efficient ones, as the source can be inferred exactly given the other parameters: there are only $n$ possible sources that we can exhaust.

The moves on the sources were almost useless given the other moves and were very seldom accepted. The moves on the parameters could have benefitted from better implementation, as they resulted in low quality estimates, although their poor identifiability makes improvements to those less interesting.

\section{Additional numerical results}

\label{sec:supp:results}





\subsection{Effect of the prior on \ourmethod}

\begin{figure}
    \centering
    \includegraphics[width=1\linewidth]{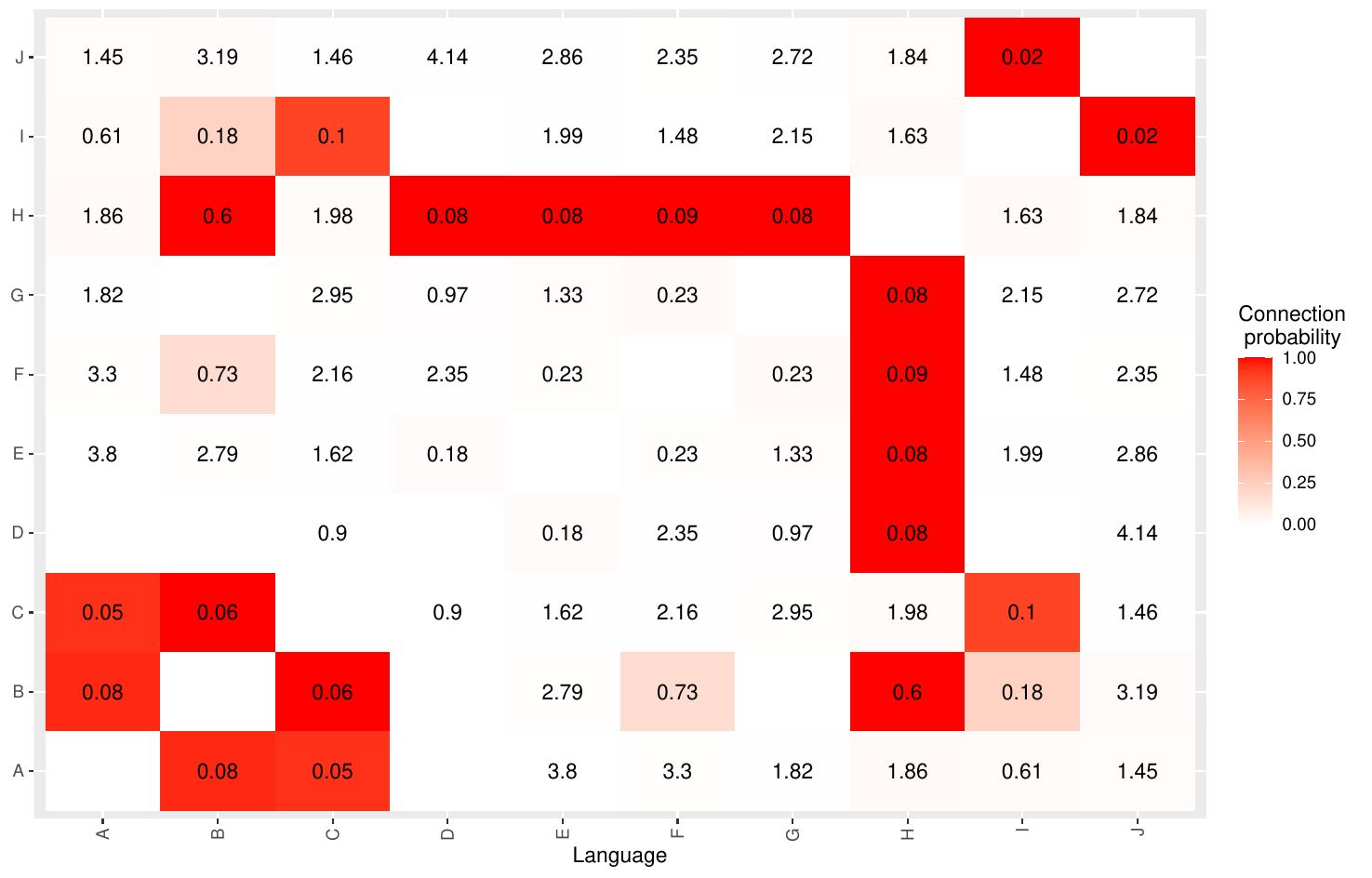}
    \caption{\ourmethod on the simulated example with the graph prior associated with $\theta=2$.}
    \label{fig:simustrongprior}
\end{figure}

First, we tried to change the parameter of the prior on the number of edges on the graph. In all the simulations presented in the main paper, $\theta=1/2$. In Figure \ref{fig:simustrongprior} we present the results. Although the chain still reconstitutes the expected groups, the mixing properties of the method are greatly decreased. This

\begin{figure}
    \centering
    \includegraphics[width=1\linewidth]{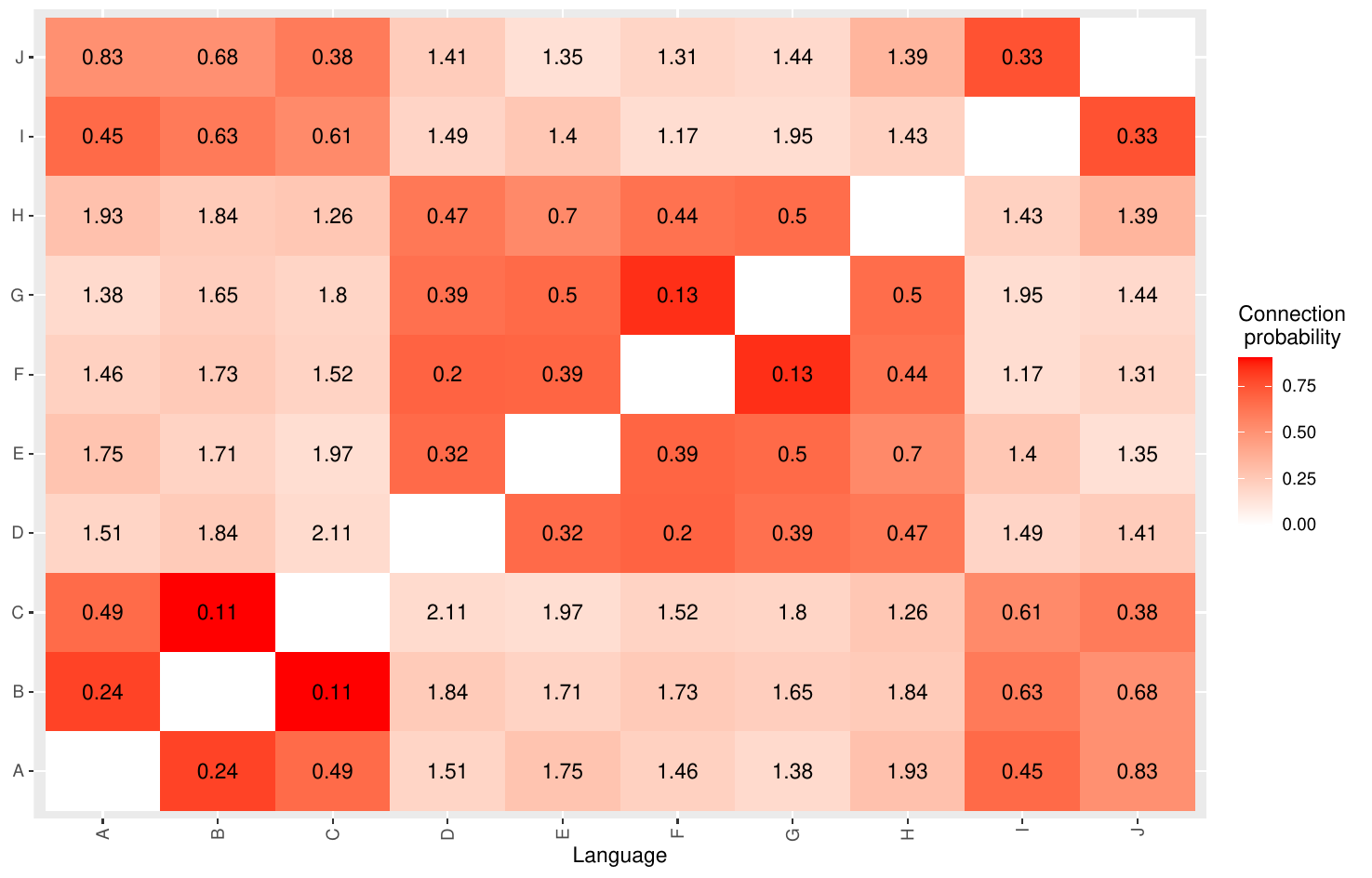}
    \caption{Results on the simulated dataset with a prior on the total length of the graph with parameter $\theta=0.5$.}
    \label{fig:simutotallength}
\end{figure}

We can also propose other types of prior on the Graph. For example, we can put a prior on the total length of the Graph: $P(E) \propto \exp(-\theta \sum_{e \in E} d_e)$. The results are similar to those presented in the main text, as can be seen in Figure \ref{fig:simutotallength}, further comforting the idea that the choice of the prior is not critical given the strength of the data signal.

\subsection{Misspecified data: multiple sources}

In order to test the resilience of \ourmethod in case of misspecifications, we tried inferring the graph on a dataset where 5\% of the traits actually come from two  sources. We can imagine an innovation that would have appeared twice and spread independently, while the data collected cannot differentiate thes innovations.

\begin{figure}
    \centering
    \includegraphics[width=\linewidth]{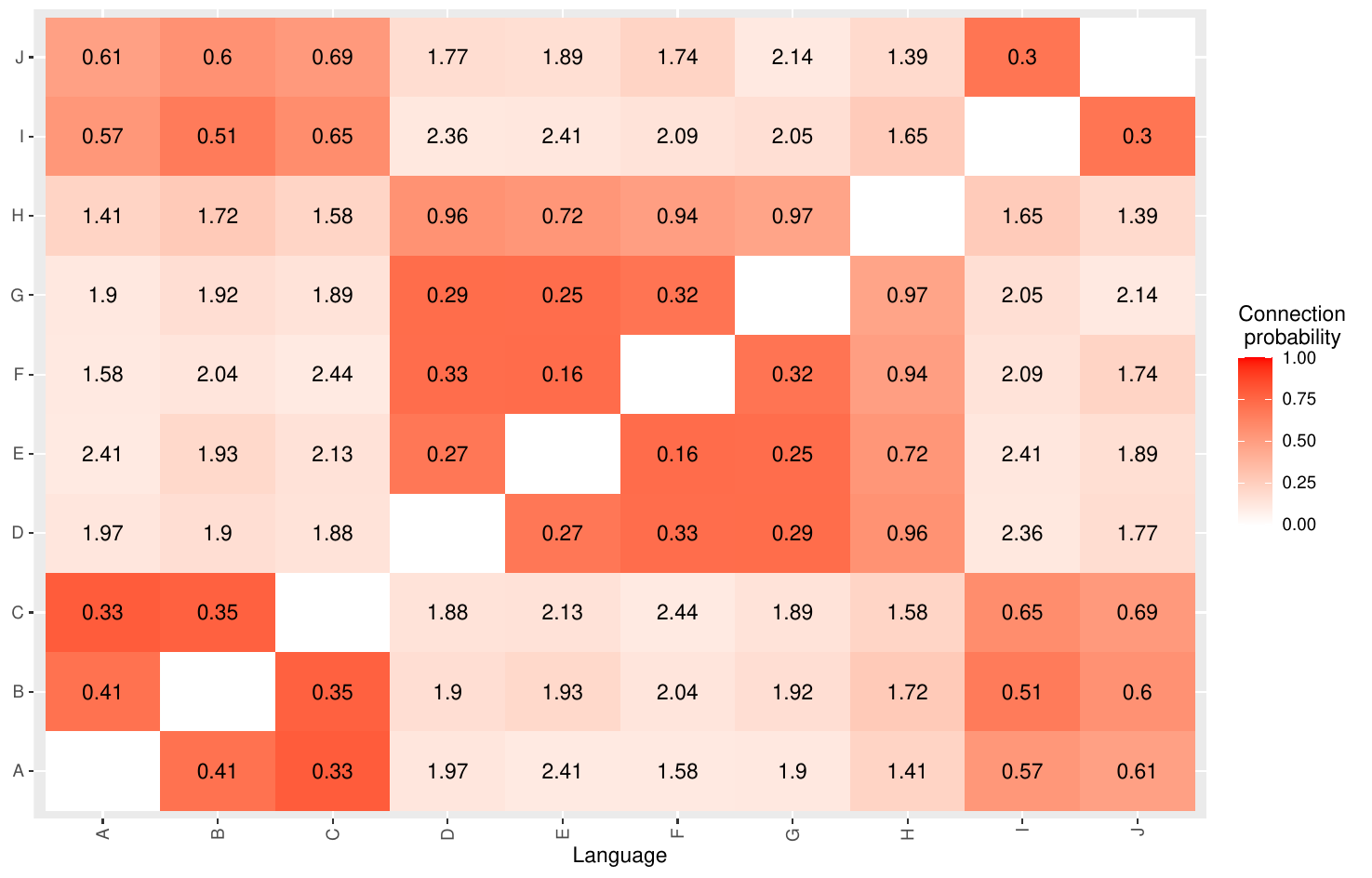}
    \caption{Result on the misspecified simulated dataset}
    \label{fig:misspec}
\end{figure}

The results presented in \ref{fig:misspec} show that the inference is not largely modified. Of course, in the case of most traits having several sources, inference is not possible.

\subsection{Another real example: Morphological data on the North Vanuatu dataset}

\begin{figure}
    \centering
    \includegraphics[width=1\linewidth]{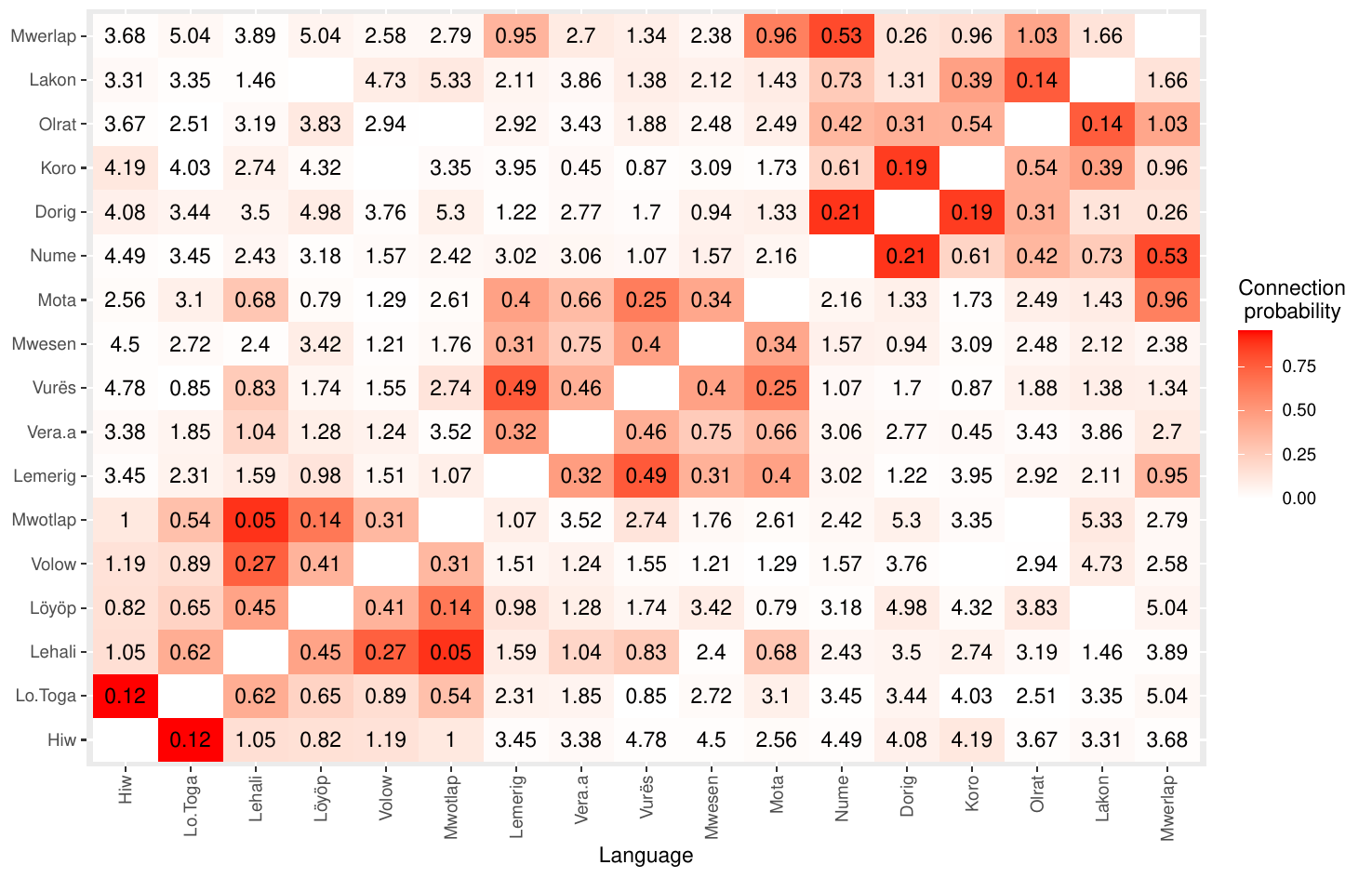}
    \includegraphics[width=1\linewidth]{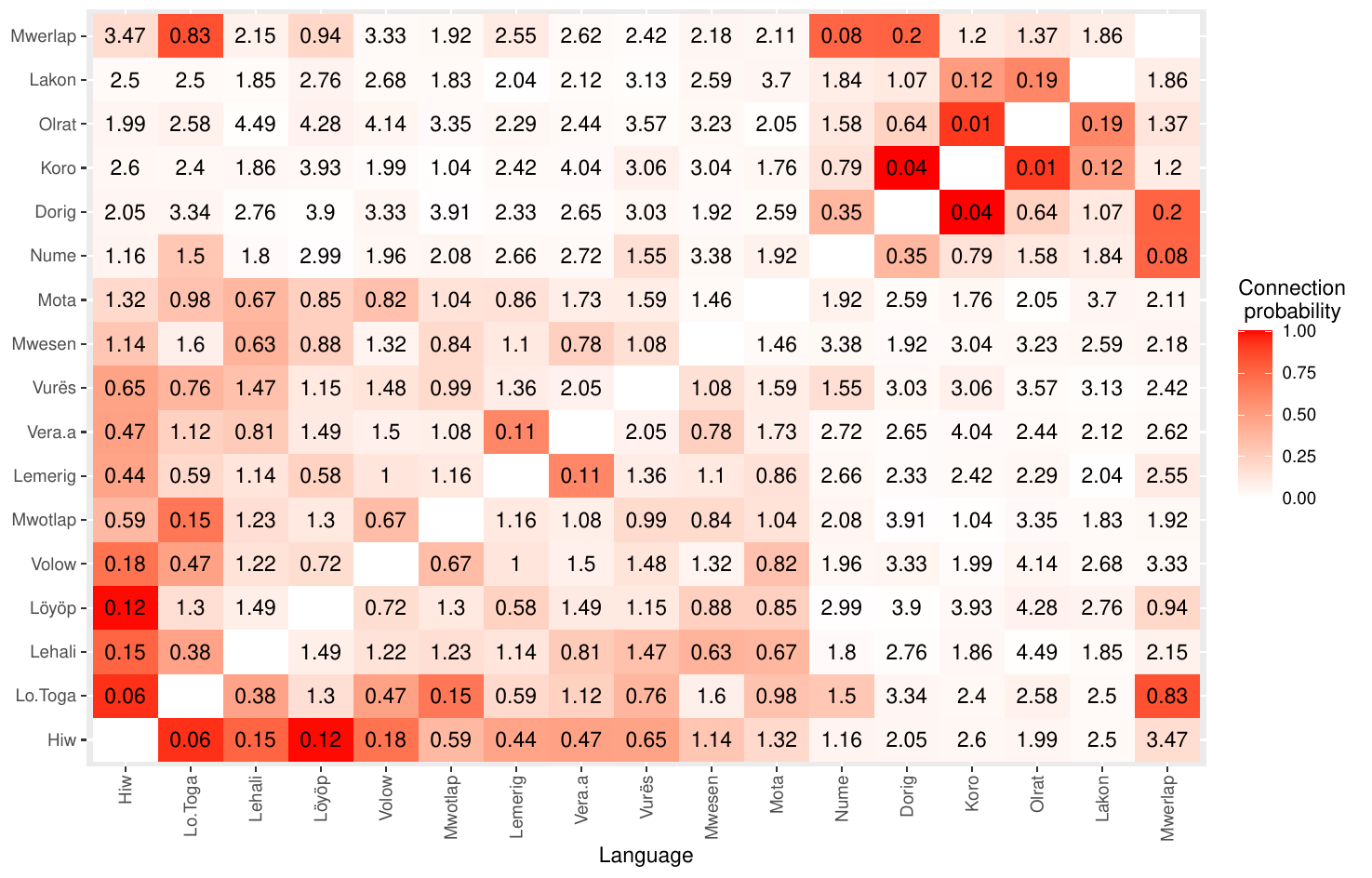}
    \caption{Two independent replicas of \ourmethod on the morphological sub dataset of the North Vanuatu dataset.}
    \label{fig:vanuatumorpho}
\end{figure}

In our examples, we used only part of the North Vanuatu dataset: lexical data. But it can also be interesting to run the same analysis with the Morphological data, which accounts for a total of 60 data lines. This number is considerably smaller than the lexical ones. We ran our method for the same number of MCMC steps and with the same parameters as the other applications. The results are presented in Fig \ref{fig:vanuatumorpho}. Overall, the results are less stable than with the lexical dataset, most likely because of the lack of data and the comparative strength of the prior. We still manage to infer two of the groups, and sometimes the three groups described previously.

These results are less interpretable than the lexical datasets, further work will be required to possibly merge morphological and lexical dataset in a single model.

\subsection{Another real example: Indo-European languages}

For this example, we used the Indo-European dataset from \url{https://github.com/evotext/ielex-data-and-tree}. As the dataset contained too many observations on too many languages (including fossiles), we reduced it to a manageable 16 languages. These were chosen by us as a set for which relations are well known and globally non debated. We selected a subset of 200 traits randomly in all the lines that were informative for our subset of languages.

\begin{figure}
    \centering
    \includegraphics[width=1\linewidth]{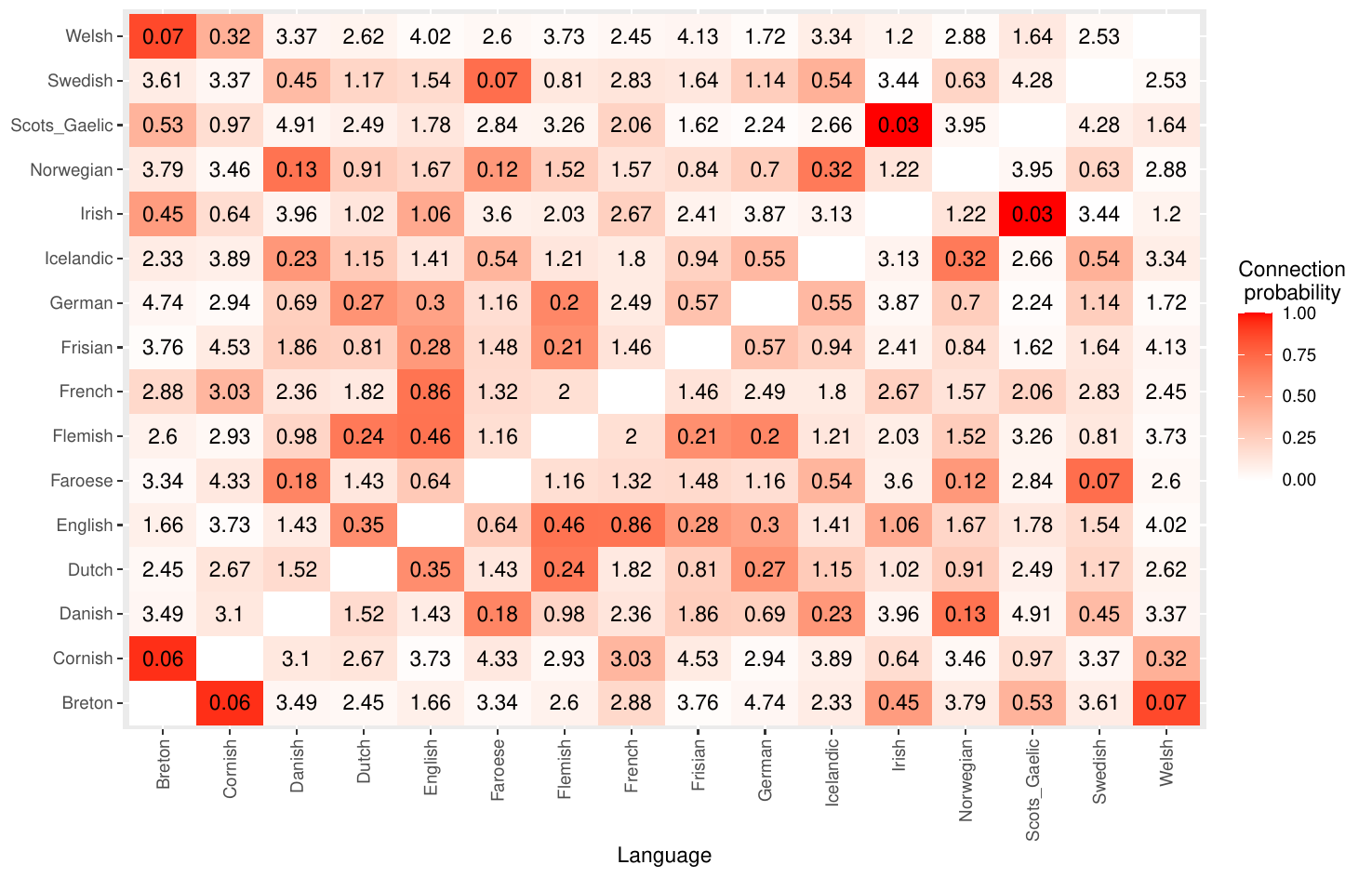}
    \includegraphics[width=1\linewidth]{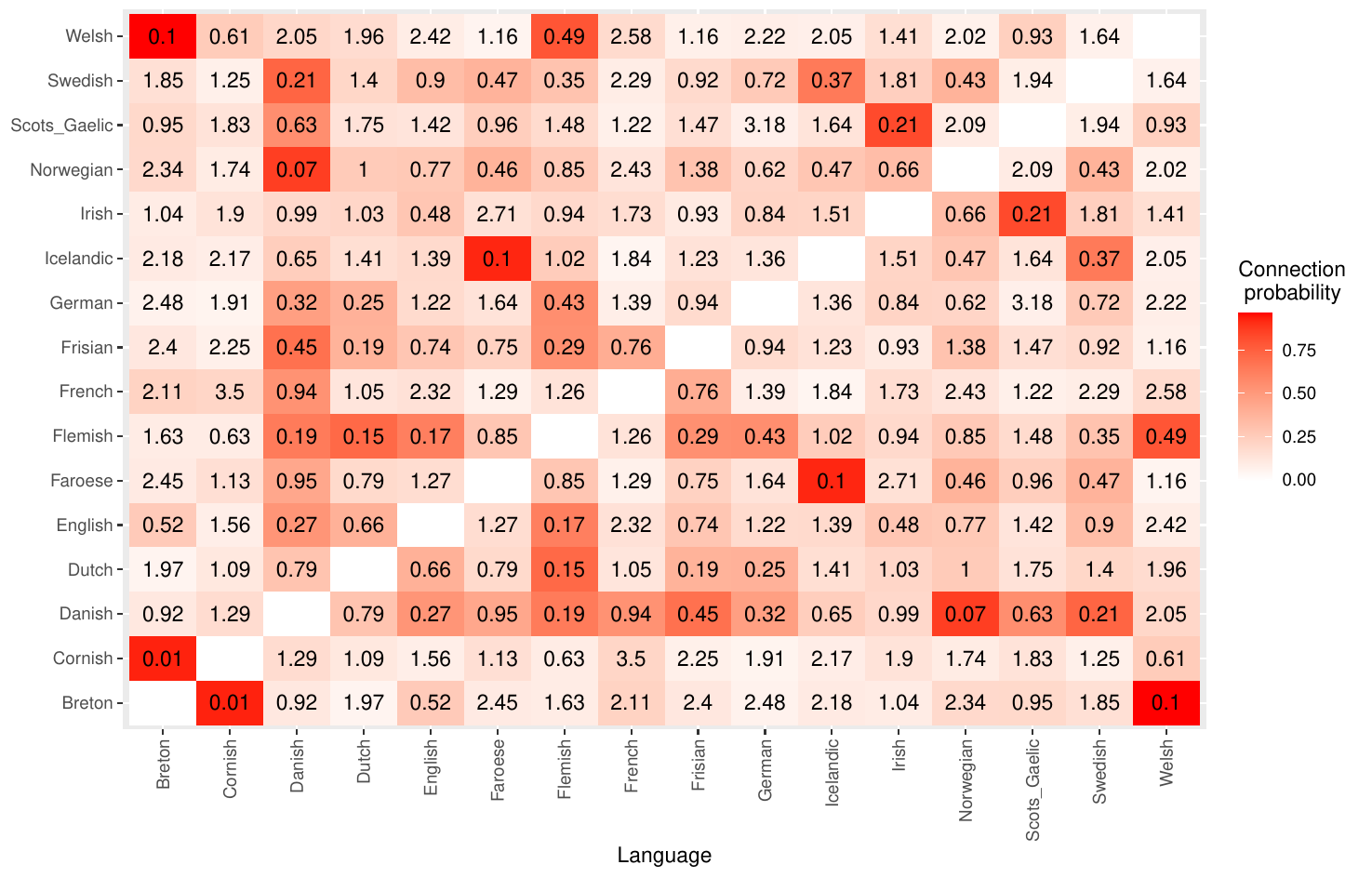}
    \caption{Two independent replicas of \ourmethod on the Indo-European dataset}
    \label{fig:IE}
\end{figure}

We ran our method for 40000 iterations with the same parameters as in the other simulations. We present the results in Figure \ref{fig:IE}. Overall, some feature appear clearly: Celtic languages are grouped together, Scandinavian languages as well, but the overall position of the languages appear unclear. This can be explained by the choice of the languages: French in our dataset is the only romance language, grouping it with any of the other languages seem uncertain, but it has to be grouped, which is why all the edge connecting french are always inferred to be relatively long, except the strange connection with Danish in the second replica.

In the second replica, the central role of Danish can be explained by a prior artifact: a star shaped graph has a lower prior than a more connected one.

\subsection{Another real example: Cultural dataset}

\begin{figure}
    \centering
    \includegraphics[width=0.5\linewidth]{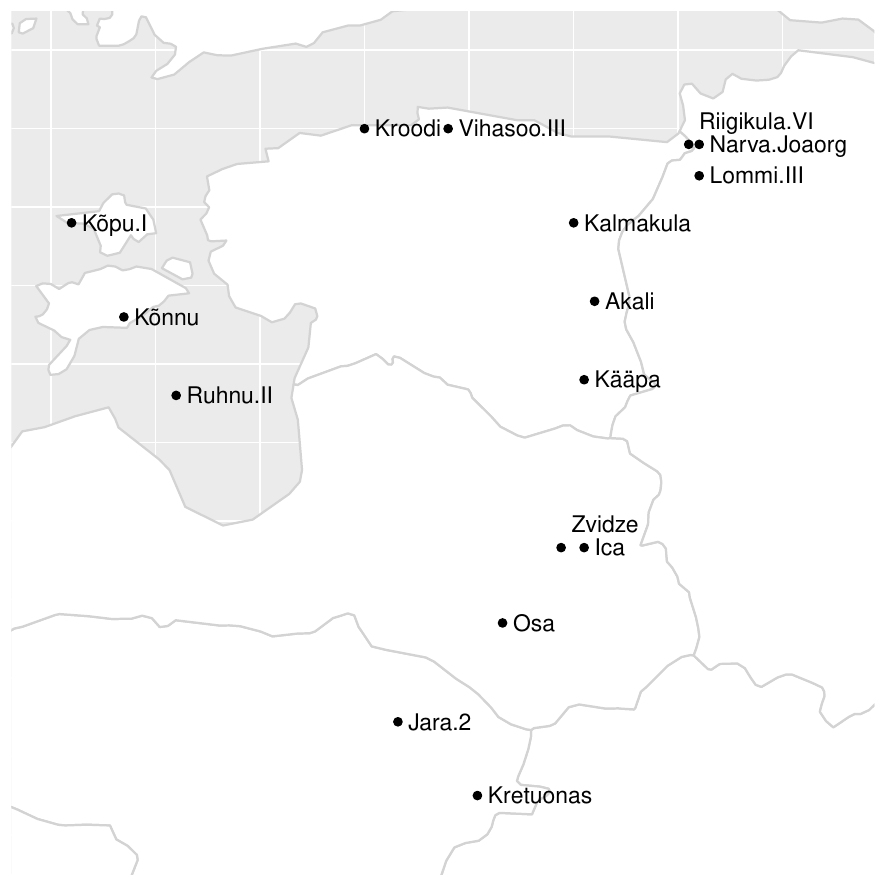}
    \caption{Sites included in the Ceramic dataset}
    \label{fig:balticceramic}
\end{figure}

In \citet{dolbunova2023transmission}, the author compile a large dataset of pottery style, morphology and techniques on different sites of prehistoric hunter-gatherers. They produce a NeighborNet and use it to infer dates on the start of pottery use. We reduced this dataset to only a subpart of the sites, in Eastern Baltic (as this dataset was adapted in size), and used \ourmethod on it. We represent in Fig. \ref{fig:balticceramic} the different sites studied.

The results are globally in line with geographical data; with a northwestern and a southeastern group. Although the sites are not perfectly synchronous, and the posterior is not very concentrated on a particular graph.

\begin{figure}
    \centering
    \includegraphics[width=1\linewidth]{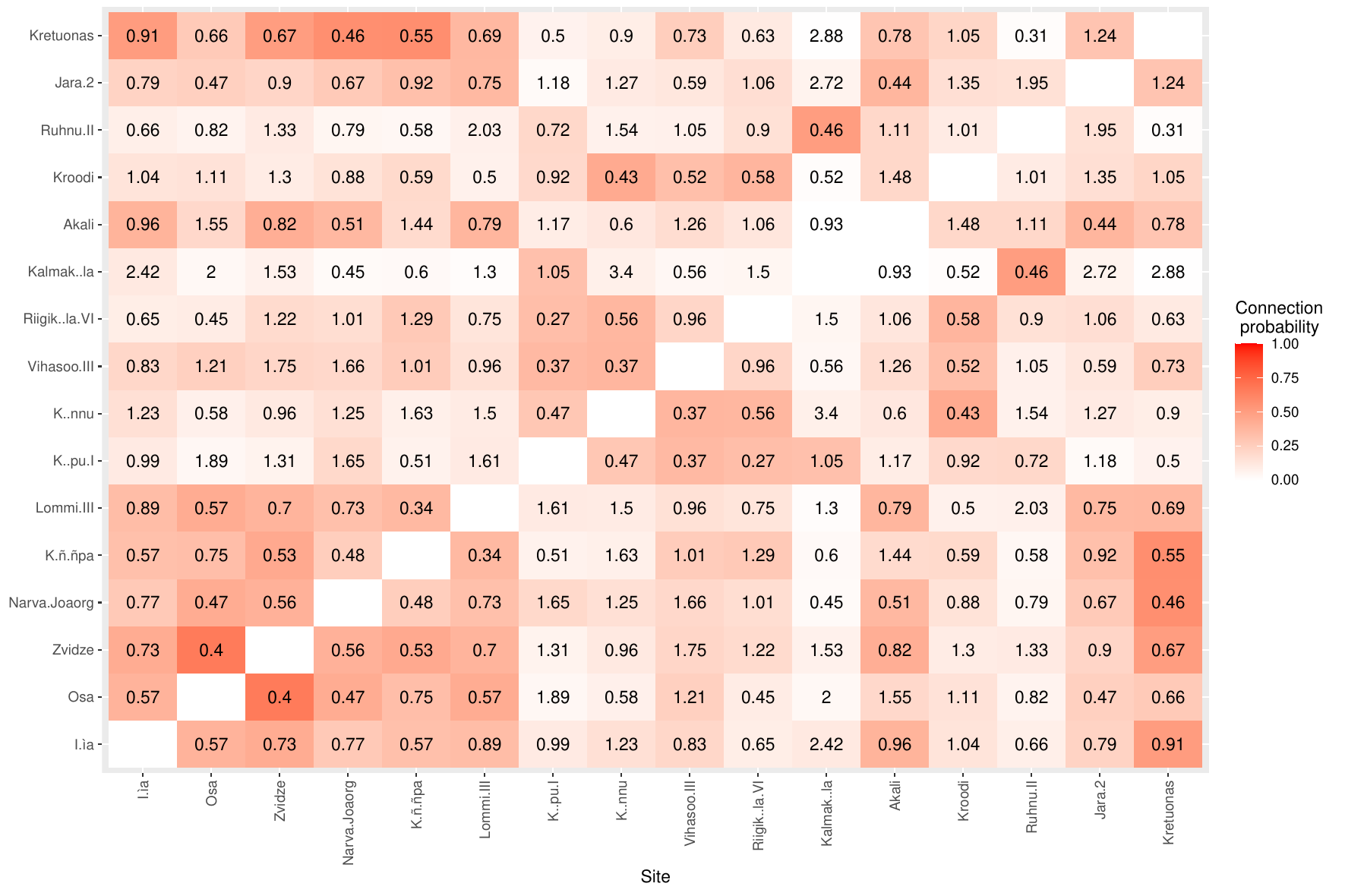}
    \includegraphics[width=1\linewidth]{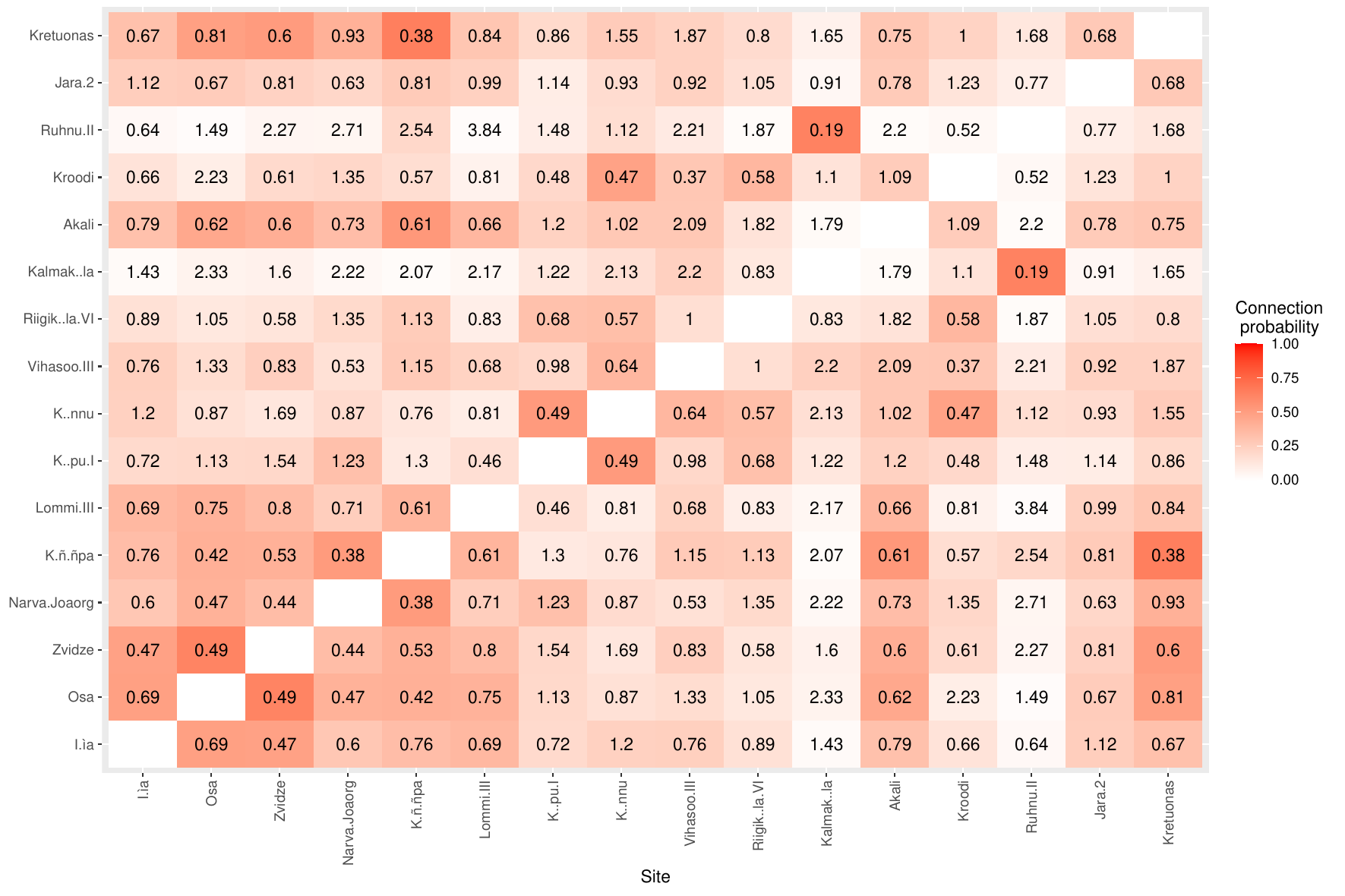}
    \caption{Results of \ourmethod on the Ceramic dataset.}
    \label{fig:ceramicresults}
\end{figure}

\begin{figure}
    \centering
    \includegraphics[width=1\linewidth]{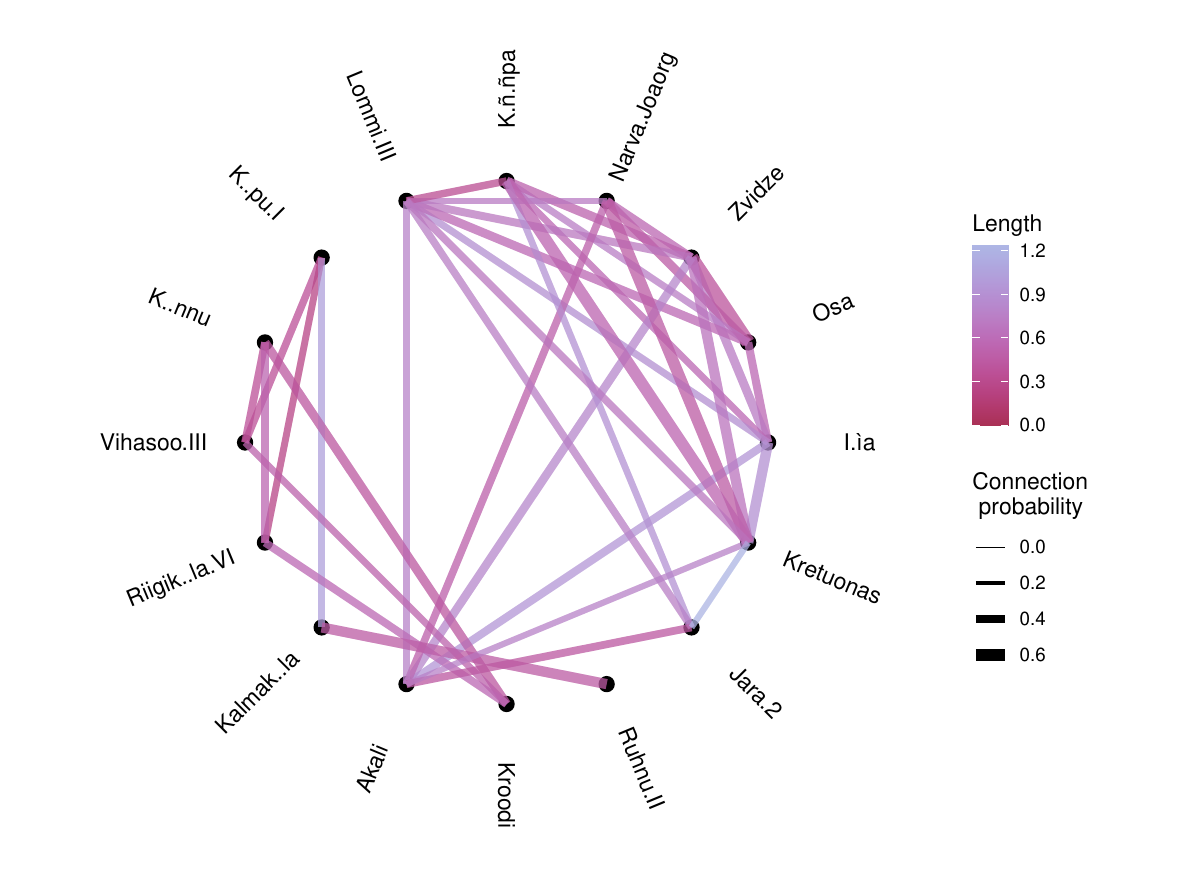}
    \includegraphics[width=1\linewidth]{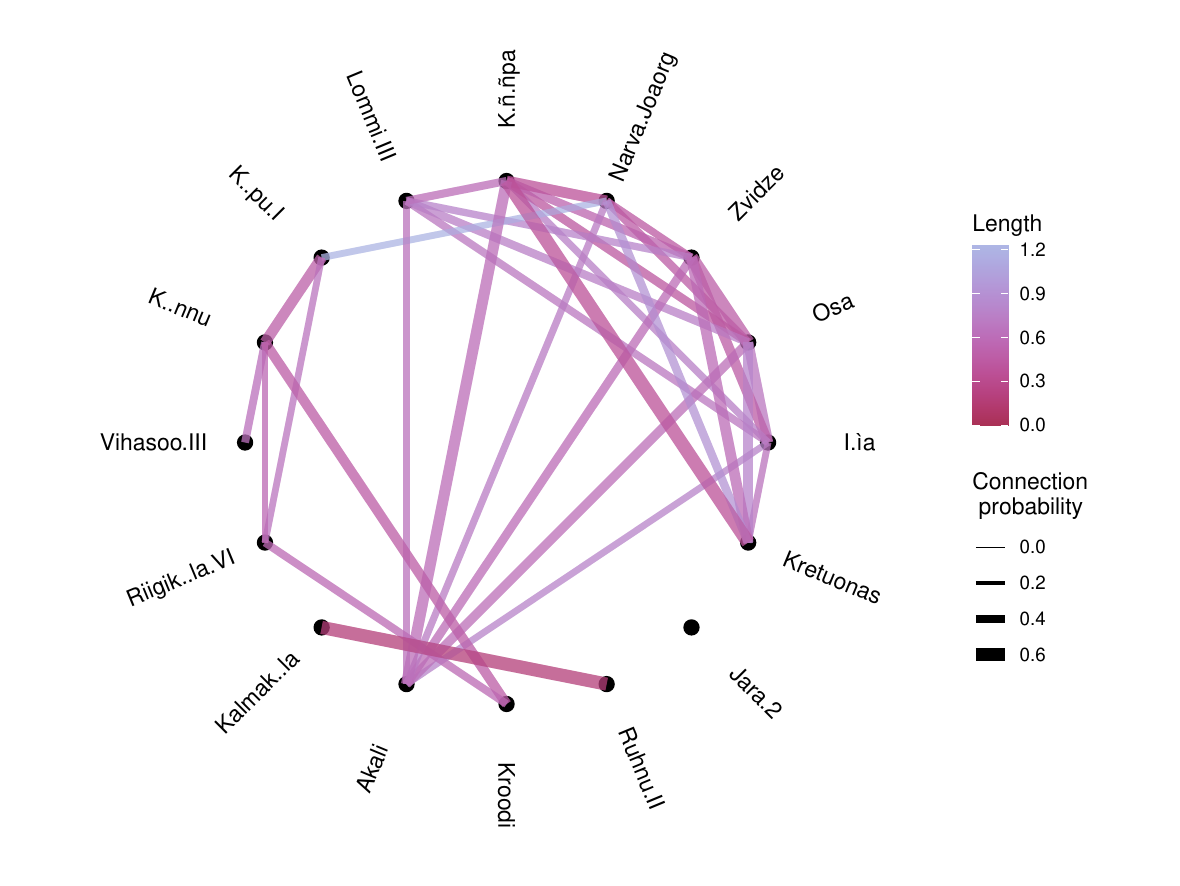}
    \caption{Results of \ourmethod on the Ceramic dataset with less than $0.3$ posterior probability removed.}
    \label{fig:ceramicresultsnetwork}
\end{figure}

\end{document}